\documentclass[preprint,12pt]{elsarticle}

\usepackage{epsfig}

\usepackage{amsthm}

\usepackage{graphicx}
\usepackage{amsmath,amstext,amsbsy,amssymb}
\usepackage[cdot,amssymb]{SIunits}
\usepackage[latin1]{inputenc}
\usepackage{float}

\begin{document}

\begin{frontmatter}



\title{Modal analysis of masonry structures}


\author{Maria Girardi}
\author{Cristina Padovani}
\author{Daniele Pellegrini}
\address{Institute of Information Science and Technologies "A. Faedo", ISTI--CNR, Via Moruzzi 1 56126 Pisa, Italy}

\begin{abstract}
This paper presents a new numerical tool for evaluating the
vibration frequencies and mode shapes of masonry buildings in the
presence of cracks. The algorithm has been implemented within the
NOSA-ITACA code, which models masonry as a nonlinear elastic
material with zero tensile strength. Some case studies are reported,
and the
differences between linear and nonlinear behaviour highlighted.\\
\end{abstract}

\begin{keyword}
masonry--like materials; modal analysis; numerical methods;
nonlinear dynamics.


\end{keyword}

\end{frontmatter}

\section{Introduction}
\label{sec1} \vspace{-2pt}

 The worldwide architectural heritage is
in pressing need of maintenance and restoration. Old structures are,
in fact, threatened by numerous environmental and anthropogenic
actions. In Italy, where seismic actions affect most of the
territory, safeguarding ancient masonry buildings and monuments is a
crucial issue for institutions and individuals alike. In this
context, the use of non--destructive techniques able to evaluate the
structural health of old buildings has a vital role to play.

Traditional building monitoring protocols generally involve visual inspections
and the measurement of some quantities (displacements, local stresses
and crack width, etc.) and their variation in time, while the
mechanical properties of the constituent materials are generally
assessed locally through destructive or non--destructive tests. More
recently ambient vibration tests, first adopted in civil
engineering to assess the structural health of bridges and tall
buildings, have assumed an increasingly important place in the fields of conservation and restoration.
This is for the most part due to the availability of very sensitive measurement
devices and ever more powerful techniques for numerical simulation.
These tests allow measuring structures' dynamic properties, such as natural frequencies, mode shapes and damping
ratios and, if coupled with a finite--element model, can provide important
information on the mechanical properties of a building's
constituent materials and boundary conditions. Moreover, long--term
monitoring protocols allow measuring any variation in the dynamic
behaviour of the monitored structures: once the influence of
environmental factors such as temperature and humidity has been
accounted for, this variation can reveal the presence of any structural damage.
Particular attention must be devoted in the monitoring of old masonry
buildings, since their make--up and behaviour are profoundly different from modern
structures. In fact,  masonry is unable to withstand tensile
stresses or large compressive stresses. For this reason, old
buildings generally present damage scenarios, with cracking induced by
the permanent loads or caused by some accidental events occurred
during their long history. The dynamic behaviour of these structures
should be analyzed taking into account the existing damage.

Numerous examples can be found in the literature of methods for detecting the
effects of damage on the dynamic properties of structures. A number of
analytical models have been formulated for predicting the
effects of cracks on the vibration frequencies of beams and rotors \cite{DIMA}.
A review of methods for detecting
structural damage by measuring the changes over time of the natural
frequencies is presented in \cite{SAL}. Changes in mode shapes are
effective damage indicators \cite{HELICOPTER}, \cite{AGAR} and
allow better localization of the damage location within the structure. In
particular, modal curvatures seem to be very sensitive to local
damage \cite{WAHAB}, though  their measurement requires the use of a
large number of sensors. Wide variations in modal parameters can be
observed when damage occurs in zones with high modal curvatures. With
regard to the monitoring of old masonry structures, a number of
papers have recently been devoted to the analysis of data
from long--term monitoring protocols; these have highlighted changes in the
modal characteristics due to environmental factors \cite{AZZARA},
\cite{CABBOI}, \cite{RAMOS}, or, in some cases, structural damage
\cite{abbiati}, \cite{LOURENCO}, \cite{saisi13}, \cite{saisi15}.

Much effort has also been devoted to numerically simulating the
effects of damage on the vibrations of the structures. In
\cite{DEROECK1} and \cite{DEROECK2} a damage identification
procedure is presented to detect and localize damage in beam
structures. A beam is discretized into finite--elements and the
damage to it simulated by reducing the stiffness matrix of some selected
elements via a damage index. The equations obtained enable obtaining
the beam's frequencies and mode shapes when the damage indices are
fixed or, viceversa, calculating the damage indices when frequencies and mode
shapes are measured on real beams. Model updating procedures are
employed in \cite{JAY} in order to fit experimental data with
finite--element models; the parameters to be updated are the Young's
moduli of the constituent materials. With regard to masonry
buildings, a common approach consists of simulating the existing
damage actually observed on the structure, by reducing the stiffness
of those elements of the finite element model belonging to the cracked
or damaged parts
 \cite{PINEDA}, \cite{LOURENCO}. In \cite{bui}
and \cite{cakti}  the dynamic properties of some masonry structures
at different damage levels are investigated via the discrete element
method.

The work presented in \cite{GIR} formulates an explicit expression linking the
fundamental frequency of a masonry--like beam to its maximum
transverse displacement. The masonry--like constitutive equation
models masonry as a non--linear elastic material with different
strengths under tension and compression \cite{GDP}, \cite{SDP}. This
constitutive equation has been implemented in the finite--element
code NOSA for the static analysis of masonry bodies \cite{SPRING}.
 The NOSA code and its
updated version NOSA--ITACA 1.0 \cite{NOSA}, have been applied to the
static and dynamic analysis of many important monuments in Italy.
NOSA--ITACA 1.0 is freely downloadable at \cite{nosaitaca}. The non--linear
equilibrium equations of masonry--like bodies are solved
via the Newton--Raphson method, by taking into account an explicit
expression of the derivative of the stress with respect to the
strain \cite{SPRING}, which allows calculation of the system's
tangent stiffness matrix.

This paper presents a numerical procedure, implemented in a newly
updated version 1.2 of the NOSA--ITACA code, which can evaluate the
natural frequencies and mode shapes of masonry buildings in the
presence of cracks. Once the initial loads and boundary conditions
have been applied to the finite--element model, the resulting
non--linear equilibrium problem is solved through an iterative
scheme. Then, a modal analysis is performed, by using the tangent
stiffness matrix calculated by the code in the last iteration before
convergence is reached. The incremental approach used by NOSA--ITACA
1.2 allows the user to perform the modal analysis at the initial
step, before application of any load to the structure, and then for
different loading steps. Following the prestressed modal analysis
applied to problems with geometric nonlinearity \cite{prestressed}
and \cite{noble}, the proposed procedure allows the user to
automatically take into account the stress distribution on the
system's stiffness matrix, thereby evaluating the effects of the
presence of cracked and crushed material on the structure's dynamic
properties.

Section \ref{sec2} briefly recalls the constitutive equation of masonry--like
materials. The explicit expressions for the
stress function and its derivative, obtained in \cite{SPRING}, are
summarized, and the numerical procedure for solving static and dynamic
problems of masonry structures, implemented in the NOSA--ITACA 1.0
code, are sketched out. Finally, the modal analysis of linear elastic
structures is addressed and the procedures implemented in the
NOSA--ITACA 1.0 code to solve the constrained generalized eigenvalue
problem resulting from application of the finite-element method is
outlined \cite{POR}. Section \ref{sec3} presents the new procedure
implemented in the NOSA--ITACA 1.2 code. The code
allows for calculating the natural frequencies and mode shapes of a
masonry body subjected to prescribed boundary conditions and loads.
The new procedure takes into account the non--linearity of the
constitutive equation of masonry--like materials as well as the
presence of cracks due to the applied loads. In general, the results
of the  modal analysis conducted after application of the
assigned loads differ from those obtained via the standard modal
analysis and can be used to assess the presence of damaged zones in
the structure. Section \ref{sec4} presents some applications of the method.
The first two cases deal with a masonry beam and an arch on piers. Both
structures are subjected to incremental loads and analyzed with the
NOSA--ITACA code. The aim is to compare the natural frequencies and
mode shapes in the linear elastic case with those calculated in the
presence of the crack distribution induced by the increasing loads.
The third case deals with an actual structure, the bell tower of
San Frediano in Lucca. The tower has been instrumented with four
high--sensitivity triaxial seismometric stations \cite{sanfrediano},
and its first five natural frequencies have been determined via OMA
techniques \cite{BRINCKER}. Then the NOSA--ITACA code, together with
model updating techniques, has been employed in order to fit the
experimental results. The model updating in the linear elastic case
is compared to the model updating applied to the tower
subjected to its own weight by taking into account the crack
distribution induced by the loads.

\section{The masonry--like constitutive equation and the NOSA-ITACA 1.0 code}
\label{sec2}

Let $Lin$ be the set of all second-order tensors with the scalar product $%
\mathbf{A}\cdot \mathbf{B}=tr(\mathbf{A}^{T}\mathbf{B})$ for any $\mathbf{A},%
\mathbf{B}\in $ $Lin$, with $\mathbf{A}^{T}$ the transpose of
$\mathbf{A}$. For $Sym$ the subspace of symmetric tensors, $Sym^{-}$
and $Sym^{+}$ are the
sets of all negative-semidefinite and positive-semidefinite elements of $Sym$%
. Given the symmetric tensors $\mathbf{A}$ and $\mathbf{B}$, we denote by $%
\mathbf{A}\otimes \mathbf{B}$ the fourth-order tensor defined by $\mathbf{A}%
\otimes \mathbf{B}[\mathbf{H}]=(\mathbf{B}\cdot \mathbf{H})\mathbf{A}$ for $%
\mathbf{H}\in Lin$ and by $\mathbb{I}_{Sym}$ the fourth-order
identity tensor on $Sym$.
For $\mathbf{a}$ and $\mathbf{b}$ vectors, the dyad $\mathbf{a}%
\otimes \mathbf{b}$ is defined by $\mathbf{a}\otimes \mathbf{bh}=(\mathbf{b}%
\cdot \mathbf{h})\mathbf{a},$ for any vector $\mathbf{h}$, with
$\cdot $ the scalar product in the space of vectors.

Let $\mathbb{C}$ be the isotropic fourth-order tensor of elastic
constants
\begin{equation}
\mathbb{C}=2\mu \mathbb{I}_{Sym}+\lambda \mathbf{I}\otimes
\mathbf{I}, \label{C}
\end{equation}%
where $\mathbf{I}\in Sym$ is the identity tensor and $\mu $ and
$\lambda $
are the Lam\'{e} moduli of the material satisfying the conditions%

\begin{equation}
\mu >0,  \textup{ \ \ }\lambda \geq 0.  \label{condC}
\end{equation}

$\mathbb{C}$ is symmetric,

\begin{equation}
\mathbf{A}\cdot \mathbb{C}[\mathbf{B}]=\mathbf{B}\cdot
\mathbb{C}[\mathbf{A}],\,\,\,\,\textup{for all}\,\,\,\,
\mathbf{A},\mathbf{B}\in Sym, \label{symmc}
\end{equation}

\noindent and in view of (\ref{condC}) is positive-definite on
$Sym$,

\begin{equation}
\mathbf{A}\cdot \mathbb{C}[\mathbf{A}]> 0\,\,\,\,\textup{for
all}\,\,\,\, \mathbf{A}\in Sym,\,\,\,\, \mathbf{A}\neq 0.
\label{posc}
\end{equation}

A masonry-like material is a nonlinear elastic material
\cite{SPRING} characterized by the fact that, for
$\mathbf{E}\in Sym$, there exists a unique triplet $(\mathbf{T},\mathbf{E}%
^{e},\mathbf{E}^{f})$ of elements of $Sym$ such that
\begin{equation}
\mathbf{E}=\mathbf{E}^{e}+\mathbf{E}^{f},  \label{EC1}
\end{equation}%
\begin{equation}
\mathbf{T}=\mathbb{C}[\mathbf{E}^{e}],  \label{EC2}
\end{equation}%
\begin{equation}
\mathbf{T}\in Sym^{-},\textup{ \ \ }\mathbf{E}^{f}\in Sym^{+},
\label{EC3}
\end{equation}%
\begin{equation}
\mathbf{T}\cdot \mathbf{E}^{f}=0.  \label{EC4}
\end{equation}%
$\mathbf{T}$ is the Cauchy stress corresponding to strain $\mathbf{E}$. Tensors $%
\mathbf{E}^{e}$ and $\mathbf{E}^{f}$ are respectively the elastic
and inelastic part of $\mathbf{E}$; $\mathbf{E}^{f}$ is also
called the fracture strain. The stress function
$\mathbb{T}:Sym\rightarrow Sym$
is given by%
\begin{equation}
\mathbb{T}(\mathbf{E})=\mathbf{T},\,\,\,\,\textup{for any \ \ }
\mathbf{E}\in Sym,\label{SE}
\end{equation}%
with $\mathbf{T}$ satisfying \eqref{EC1}--\eqref{EC4}.
The explicit expression for the stress function $%
\mathbb{T}$, calculated in \cite{SPRING}, is recalled in the
following.

For $\mathbf{E}\in Sym,$ let $e_{1}\leq e_{2}\leq e_{3}$ be its
ordered eigenvalues and $\mathbf{q}_{1},$
$\mathbf{q}_{2},\mathbf{q}_{3}$ the corresponding eigenvectors. We
introduce the orthonormal basis of $Sym$
(with respect to the scalar product $\cdot $)%
\begin{equation*}
\mathbf{O}_{11}=\mathbf{q}_{1}\otimes \mathbf{q}_{1},\textup{ }\mathbf{O}_{22}=%
\mathbf{q}_{2}\otimes \mathbf{q}_{2},\textup{ }\mathbf{O}_{33}=\mathbf{q}%
_{3}\otimes \mathbf{q}_{3},\textup{ }
\end{equation*}%
\begin{equation}
\mathbf{O}_{12}=\frac{1}{\sqrt{2}}(\mathbf{q}_{1}\otimes \mathbf{q}_{2}+%
\mathbf{q}_{2}\otimes \mathbf{q}_{1}),\textup{ }\mathbf{O}_{13}=\frac{1}{\sqrt{%
2}}(\mathbf{q}_{1}\otimes \mathbf{q}_{3}+\mathbf{q}_{3}\otimes \mathbf{q}%
_{1}),
\end{equation}%
\begin{equation*}
\mathbf{O}_{23}=\frac{1}{\sqrt{2}}(\mathbf{q}_{2}\otimes \mathbf{q}_{3}+%
\mathbf{q}_{3}\otimes \mathbf{q}_{2}).
\end{equation*}%

\vspace{6pt} \noindent

Given $\mathbf{E}$, the corresponding stress $\mathbf{T}$ satisfying
the constitutive equation of masonry-like materials
\eqref{EC1}--\eqref{EC4} is given by
\begin{equation}
\textup{if }\mathbf{E}\in V_{0}\textup{ then }\mathbf{T}=\mathbf{0},
\label{T0}
\end{equation}%

\begin{equation}
\textup{if }\mathbf{E}\in V_{1}\textup{ then
}\mathbf{T}=Ee_{1}\mathbf{O}_{11}, \label{T1}
\end{equation}%

\begin{equation*}
\textup{if }\mathbf{E}\in V_{2} \textup{ then }\mathbf{T}=\frac{2\mu }{2+\alpha }%
\left\{ [2(1+\alpha )e_{1}+\alpha e_{2}]\mathbf{O}_{11}\right.
\end{equation*}%
\begin{equation}
\left. +[\alpha e_{1}+2(1+\alpha )e_{2}]\mathbf{O}_{22}\right\},
\label{T2}
\end{equation}%

\begin{equation}
\textup{if }\mathbf{E}\in V_{3}\textup{ then
}\mathbf{T}=\mathbb{C}[\mathbf{E}], \label{T3}
\end{equation}%
where the sets $V_{k}$ are%
\begin{equation}
V_{0}=\left\{ \mathbf{E}\in Sym\textup{ : }e_{1}\geq 0\right\} ,
\label{V0}
\end{equation}%

\begin{equation}
V_{1}=\left\{ \mathbf{E}\in Sym\textup{ : }e_{1}\leq 0,\textup{
}\alpha e_{1}+2(1+\alpha )e_{2}\geq 0\right\} ,  \label{V1}
\end{equation}%

\begin{equation}
V_{2}=\left\{ \mathbf{E}\in Sym\textup{ : }\alpha e_{1}+2(1+\alpha
)e_{2}\leq 0,\textup{ }2e_{3}+\alpha tr\mathbf{E}\geq 0\right\} ,
\label{V2}
\end{equation}%

\begin{equation}
V_{3}=\left\{ \mathbf{E}\in Sym\textup{ : }2e_{3}+\alpha
tr\mathbf{E}\leq 0\right\} ,  \label{V3}
\end{equation}%

\vspace{6pt} \noindent with $\alpha =\lambda /\mu $ and $E=\mu (2\mu
+3\lambda )/(\mu +\lambda )$ the Young's modulus.

As for the fracture strain,

\begin{equation}
\textup{if }\mathbf{E}\in V_{0}\textup{ then
}\mathbf{E}^{f}=\mathbf{E}, \label{Ef0}
\end{equation}%

\begin{equation}
\textup{if }\mathbf{E}\in V_{1}\textup{ then
}\mathbf{E}^{f}=(e_{2}+\frac{\alpha
}{2(1+\alpha )}e_{1})\mathbf{O}_{22}+(e_{3}+\frac{\alpha }{2(1+\alpha )}%
e_{1})\mathbf{O}_{33},  \label{Ef1}
\end{equation}%

\begin{equation}
\textup{if }\mathbf{E}\in V_{2}\textup{ then
}\mathbf{E}^{f}=[e_{3}+\frac{\alpha }{2+\alpha
}(e_{1}+e_{2})]\mathbf{O}_{33},  \label{Ef2}
\end{equation}%

\begin{equation}
\textup{if }\mathbf{E}\in V_{3}\textup{ then
}\mathbf{E}^{f}=\mathbf{0}. \label{Ef3}
\end{equation}
\vspace{6pt}

For $W_{k}$ the interior of $V_{k}$, function $\mathbb{T}$ turns out
to be differentiable in $W=\bigcup\limits_{i=0}^{3}W_{i}$
\cite{SPRING},\cite{PS}. Thus, for $\mathbf{E} \in W$ denoting by
$D_{E}\mathbb{T}(\mathbf{E})$ the derivative of
$\mathbb{T}(\mathbf{E})$ with respect  to $\mathbf{E}$, we have

\begin{equation}
\mathbb{T}(\mathbf{E}+\mathbf{H})=\mathbb{T}(\mathbf{E})+D_{E}\mathbb{T}(\mathbf{E})[%
\mathbf{H}]+o(\mathbf{H}),\text{ \ \ }\mathbf{H}\in Sym,\text{ \ \ }\mathbf{H%
}\rightarrow \mathbf{0}.  \label{devel}
\end{equation}
$D_{E}\mathbb{T}(\mathbf{%
E})$ is a symmetric fourth--order tensor from $Sym$ into itself and
has the following expressions in the regions $W_{i}$  \cite{SPRING}.

\begin{equation}
\textup{If }\mathbf{E}\in W_{0}\textup{ then
}D_{E}\mathbb{T}(\mathbf{E})=\mathbb{O},  \label{DT0}
\end{equation}%
where $\mathbb{O}$ is the null fourth-order tensor,
\begin{equation*}
\textup{if }\mathbf{E}\in W_{1}\textup{ then }D_{E}\mathbb{T}(\mathbf{E})=E\left( \mathbf{O}_{11}\otimes \mathbf{O}_{11}-\frac{e_{1}}{%
e_{2}-e_{1}}\mathbf{O}_{12}\otimes \mathbf{O}_{12}\right.
\end{equation*}%

\begin{equation}
\left. -\frac{e_{1}}{e_{3}-e_{1}}\mathbf{O}_{13}\otimes \mathbf{O}%
_{13}\right) ,  \label{DT1}
\end{equation}%

\begin{equation*}
\textup{if }\mathbf{E}\in W_{2}\textup{ then
}D_{E}\mathbb{T}(\mathbf{E})=2\mu \mathbf{O}_{12}\otimes
\mathbf{O}_{12}
\end{equation*}%

\begin{equation*}
-\frac{2\mu }{2+\alpha }\frac{2(1+\alpha )e_{1}+\alpha e_{2}}{e_{3}-e_{1}}%
\mathbf{O}_{13}\otimes \mathbf{O}_{13}
\end{equation*}%

\begin{equation*}
-\frac{2\mu }{2+\alpha }\frac{\alpha e_{1}+2(1+\alpha )e_{2}}{e_{3}-e_{2}}%
\mathbf{O}_{23}\otimes \mathbf{O}_{23}
\end{equation*}%

\begin{equation*}
+\frac{2\mu (2+3\alpha )}{2+\alpha }\frac{\mathbf{O}_{11}+\mathbf{O}_{22}}{%
\sqrt{2}}\otimes \frac{\mathbf{O}_{11}+\mathbf{O}_{22}}{\sqrt{2}}
\end{equation*}%
\begin{equation}
+2\mu \frac{\mathbf{O}_{11}-\mathbf{O}_{22}}{\sqrt{2}}\otimes \frac{\mathbf{O%
}_{11}-\mathbf{O}_{22}}{\sqrt{2}},  \label{DT2}
\end{equation}%

\begin{equation}
\textup{if }\mathbf{E}\in W_{3}\textup{ then
}D_{E}\mathbb{T}(\mathbf{E})=\mathbb{C} .  \label{DT3a}
\end{equation}

\vspace{6pt}

In order to study real problems, the equilibrium problem of masonry
structures can be solved via the finite element method. To this end,
suitable numerical techniques have been developed \cite{SPRING}
based on the Newton-Raphson method for solving the nonlinear system
obtained by discretising the structure into finite elements. Their
application is based on the explicit expression for the derivative
of the stress with respect to the strain, which is needed in order to
calculate the tangent stiffness matrix. The numerical method studied
and the constitutive equation of masonry-like materials described
above have therefore been implemented into the finite element code
NOSA \cite{SPRING}.

As far as the numerical solution of dynamic problems is concerned,
direct integration of the equations of motion is required
\cite{DPP}. In fact, due to the nonlinearity of the adopted
constitutive equation, the mode-superposition method is meaningless.
With the aim of solving such problems, we have instead implemented
the Newmark method \cite{bw} within NOSA in order to perform the
integration with respect to time of the system of ordinary
differential equations obtained by discretising the structure into
finite elements. Moreover, the Newton-Raphson scheme, needed to
solve the nonlinear algebraic system obtained at each time step, has
been adapted to the dynamic case. The NOSA code, devoted to static
and dynamic analyses of ancient masonry constructions, enables
determining the stress state and the presence of any cracking.
Moreover, the effects of thermal variations and the effectiveness of
various strengthening interventions, such as the application of tie
rods and retaining structures, can be evaluated, and those with the
least environmental and visual impact identified.

The code has been successfully applied to the analysis of arches and
vaults \cite{Volte}, and in a number of studies of buildings of
historical and architectural interest \cite{SPRING}, \cite{Buti}.

Within the framework of the project "Tools for modelling and
assessing the structural behaviour of ancient constructions"
\cite{nosaitaca} funded by the Region of Tuscany (2011-2013), the
NOSA code has been integrated in the open source graphic platform
SALOME \cite{salome}, used both to define the geometry of the
structure under examination and to visualise the results of the
structural analysis. The result of this integration is the freeware
code NOSA-ITACA 1.0 \cite{NOSA}, \cite{nosaitaca}. NOSA-ITACA 1.0
has been applied to several studies commissioned by both private and
public bodies of the static and dynamic behaviour of historical
masonry buildings \cite{Voltone}. In many cases, such studies have
also provided important information on the structure's seismic
vulnerability, which can be assessed with respect to current Italian
and European regulations \cite{reg}.

An efficient implementation of the numerical methods for constrained
eigenvalue problems for  modal analysis of linear elastic
structures has been embedded in NOSA--ITACA 1.0 \cite{POR}. Such
implementation takes into account both the sparsity of the matrices
and the features of master-slave constraints (tying or multipoint
constraints). It is based on the open-source packages SPARSKIT
\cite{Saad}, for managing matrices in sparse format (storage,
matrix-vector products), and ARPACK \cite{Lehoucq}, which implements
a method based on Lanczos factorization combined with spectral
techniques that improve convergence. In particular in \cite{POR} the
authors have implemented a procedure for solving the constrained
generalized eigenvalue problem
\begin{equation}\label{gep_MK}
K\, \phi = \omega^2\, M\,\phi,
\end{equation}
\begin{equation}\label{gep_MS}
 T \phi=0,
\end{equation}
with $T\in \mathbb{R}^{m\times n}$ and $m\ll n$. Equation
(\ref{gep_MK}) is derived from the equation

\begin{equation}\label{edo}
 M \ddot{u} + K u = 0,
\end{equation}
governing the free vibrations of a linear elastic structure
discretized into finite elements. In equation (\ref{edo}) $u$ is the
displacement vector, which belongs to $\mathbb{R}^{n}$ and depends
on time $t$, $\ddot{u}$ is the second-derivative of $u$ with respect
to $t$, and $K$ and $M\in \mathbb{R}^{n\times n}$ are the stiffness
and mass matrices of the finite-element assemblage. $K$ is symmetric
and positive-semidefinite, $M$ is symmetric and positive-definite,
and both are banded with bandwidth depending on the numbering of the
finite-element nodal points. Displacements $u_i$ are also called
degrees of freedom; the integer $n$ is the total number of degrees
of freedom of the system and is generally very large, since it
depends on the level of discretization of the problem. By assuming
that
\begin{equation}\label{Usin}
 u = \phi\sin(\omega t),
\end{equation}
with $\phi$ a vector of $\mathbb{R}^{n}$ and $\omega$ a real scalar,
and applying the modal superposition \cite{bw}, equation (\ref{edo})
is transformed into the generalized eigenvalue problem
(\ref{gep_MK}). Condition (\ref{gep_MS}) expresses the fixed
constraints and the master-slave relations assigned to displacement
$u$, written in terms of vector $\phi$. The restriction of the
matrix $K$ to the null subspace of $\mathbb{R}^{n}$ defined by
(\ref{gep_MS}) is positive-definite.

Although the constitutive equation \eqref{EC1}--\eqref{EC4} adopted
for masonry is nonlinear, the modal analysis, which is based on the
assumption that the materials constituting the construction are
linear elastic, is widely used in applications and furnishes
important qualitative information on the dynamic behavior of masonry
structures and thereby allows for assessing their seismic
vulnerability in light of Italian  and European regulations
\cite{reg}. On the other hand, traditional modal analysis does
not take into account the influence that both the nonlinear
behaviour of the masonry material and the presence of cracked
regions can have on the natural frequencies of masonry structures.
The approach described in the next section and implemented in NOSA--ITACA 1.2 allows for calculating
the natural frequencies and modal shapes of a masonry body subjected to
given system of loads and exhibiting a crack distribution due to
these loads.

\section{A new procedure for the modal analysis of masonry--like structures} \label{sec3}
Let us consider a body\footnote{A body is a regular region of the
three-dimensional Euclidean space having boundary
$\partial\mathcal{B}$ , with outward unit normal $\mathbf{n}$
\cite{Gurtin2}.} $\mathcal{B}$ whose boundary $\partial\mathcal{B}$
is composed of two complementary and disjoined portions
$\partial\mathcal{B}_1$ and $\partial\mathcal{B}_2$.  $\mathcal{B}$
is made of a masonry-like material with constitutive equation
\eqref{EC1}--\eqref{EC4} and mass density $\rho$. Given the loads
($\mathbf{b}$, $\mathbf{s_0}$), with the body force $\mathbf{b}$
defined over $\mathcal{B}$ and the surface force $\mathbf{s_0}$
defined over $\partial\mathcal{B}_2$, let ($\mathbf{\widetilde{u}}$,
$\mathbf{\widetilde{E}}$, $\mathbf{\widetilde{T}}$) a triple
consisting of one vector and two tensor fields defined over
$\mathcal{B}$ satisfying the following equilibrium problem
\cite{SPRING}

\begin{equation}
div\mathbf{\widetilde{T}}+\mathbf{b}=\mathbf{0}
\,\,\,\,\,\,\text{on}\,\, \mathcal{B},  \label{eq1}
\end{equation}

\begin{equation}
\mathbf{\widetilde{E}}=\frac{\nabla \mathbf{\widetilde{u}}+\nabla
\mathbf{\widetilde{u}}^{T}}{2} \,\,\,\,\,\,\text{on}\,\,
\mathcal{B},  \label{con1}
\end{equation}

\begin{equation}
\mathbf{\widetilde{T}}=\mathbb{T}(\mathbf{\widetilde{E}})
\,\,\,\,\,\,\text{on}\,\, \mathcal{B},  \label{ec1}
\end{equation}

\begin{equation}
\mathbf{\widetilde{u}}=\mathbf{0} \,\,\,\,\,\,\text{on}\,\,
\partial\mathcal{B}_1,  \label{bc1}
\end{equation}

\begin{equation}
\mathbf{\widetilde{T}}\mathbf{n}=\mathbf{s_0}
\,\,\,\,\,\,\text{on}\,\,
\partial\mathcal{B}_2.  \label{bc2}
\end{equation}

The equilibrium problem \eqref{eq1}-\eqref{bc2} is dealt with in
\cite{SPRING}, where the uniqueness of its solution in terms of
stress is proved and the iterative procedure implemented in the NOSA
code to calculate a numerical solution is described in detail.

Let $(0,t_0)$ denote a fixed interval of time; a motion of
$\mathcal{B}$ is a vector field $\mathbf{u}$ defined on
$\mathcal{B}\times (0, t_0)$. The vector $\mathbf{u}(\mathbf{x},t)$
is the displacement of $\mathbf{x}$ at time $t$ and $\frac{\partial
^{2}\mathbf{u}}{\partial t^{2}}$ its acceleration. Let us consider
small motions $\mathbf{u}$ such that their gradient $\nabla
\mathbf{u}$ is small, and $\mathbf{u}=\mathbf{0}$ on
$\partial\mathcal{B}_1$.

For the displacement field $\mathbf{\widetilde{u}}+\mathbf{u}$, the
strain field $\mathbf{\widetilde{E}}+\mathbf{E}$, with
$\mathbf{E}=({\nabla \mathbf{u}+\nabla \mathbf{u}^{T}})/2$ and the
stress field $\mathbb{T}(\mathbf{\widetilde{E}}+\mathbf{E})$, the
equation of motion reads

 \begin{equation}
div{\large (}\mathbb{T}(\mathbf{\widetilde{E}}+\mathbf{E}){\large
)}+\mathbf{b}=\rho \frac{\partial
^{2}(\mathbf{\widetilde{u}}+\mathbf{u})}{\partial t^{2}}
\,\,\,\,\,\,\text{on}\,\, \mathcal{B}.\label{eom}
\end{equation}%
From \eqref{eom}, in view of \eqref{devel} and \eqref{ec1}, bearing
in mind the equilibrium equation \eqref{eq1} and neglecting terms of
order $o(\nabla \mathbf{u})$, we get the linearized equation of
motion

\begin{equation}
div{\large (}D_{E}\mathbb{T}(\mathbf{\widetilde{E}})[\frac{\nabla
\mathbf{u+}\nabla \mathbf{u}^{T}}{2}]{\large )}=\rho \frac{\partial
^{2}\mathbf{u}}{\partial t^{2}} \,\,\,\,\,\,\text{on}\,\,
\mathcal{B},\label{eomc}
\end{equation}%
where the fourth-order tensor $\large
D_{E}\mathbb{T}(\mathbf{\widetilde{E}})$  depends on $\mathbf{x} \in
\mathcal{B}$.

Equation \eqref{eomc}, governing the undamped free vibrations of
$\mathcal{B}$ about the equilibrium state ($\mathbf{\widetilde{u}}$,
$\mathbf{\widetilde{E}}$, $\mathbf{\widetilde{T}}$), is linear and
via the application of the finite element method can be transformed
into the equation

\begin{equation}\label{edotilde}
 M \ddot{u} + \widetilde{K} u = 0,
\end{equation}
which is analogous to  \eqref{edo}, though here the elastic
stiffness matrix $K$, calculated using the elastic tensor
$\mathbb{C}$, has been replaced by the damaged stiffness matrix
$\widetilde{K}$, calculated using $\large
D_{E}\mathbb{T}(\mathbf{\widetilde{E}})$, which takes into account
the presence of cracks in body $\mathcal{B}$.

Thus, a new numerical method for the modal analysis of masonry
constructions has been implemented in an updated version 1.2 of the NOSA-ITACA code. Given the
structure under examination, discretized into finite elements, and given
the mechanical properties of the masonry-like material constituting
the structure, together with the kinematic constraints and loads acting on
it, the procedure consists of the following steps.

Step 1. A preliminary modal analysis is conducted by assuming the
the structure's constituent material to be linear elastic. The
generalized eigenvalue problem (\ref{gep_MK})-(\ref{gep_MS}) is then
solved and the natural frequencies
$\text{f}_\text{i}^l=\omega_\text{i}^l/2\pi$ and mode shapes
$\phi_\text{i}^l$ calculated.

Step 2. The equilibrium problem (\ref{eq1})-(\ref{bc2}) is solved
and its solution, ($\mathbf{\widetilde{u}}$,
$\mathbf{\widetilde{E}}$, $\mathbf{\widetilde{T}}$), calculated. The
fourth-order tensor $\large D_{E}\mathbb{T}(\mathbf{\widetilde{E}})$
needed to calculate the damaged stiffness matrix $\widetilde{K}$ to
be used in the next step is evaluated by using
\eqref{DT0}--\eqref{DT3a}.

Step 3. The  generalized eigenvalue problem
(\ref{gep_MK})-(\ref{gep_MS}), with matrix $\widetilde{K}$ in
place of the elastic stiffness matrix $K$, is solved and the
natural frequencies $\text{f}_\text{i}=\omega_\text{i}/2\pi$ and mode shapes
$\phi_\text{i}$ of the damaged structure calculated.

\section{Case studies} \label{sec4}

\subsection{The masonry beam} \label{sec4_1}
\begin{figure}\centering
\includegraphics[width=15cm]{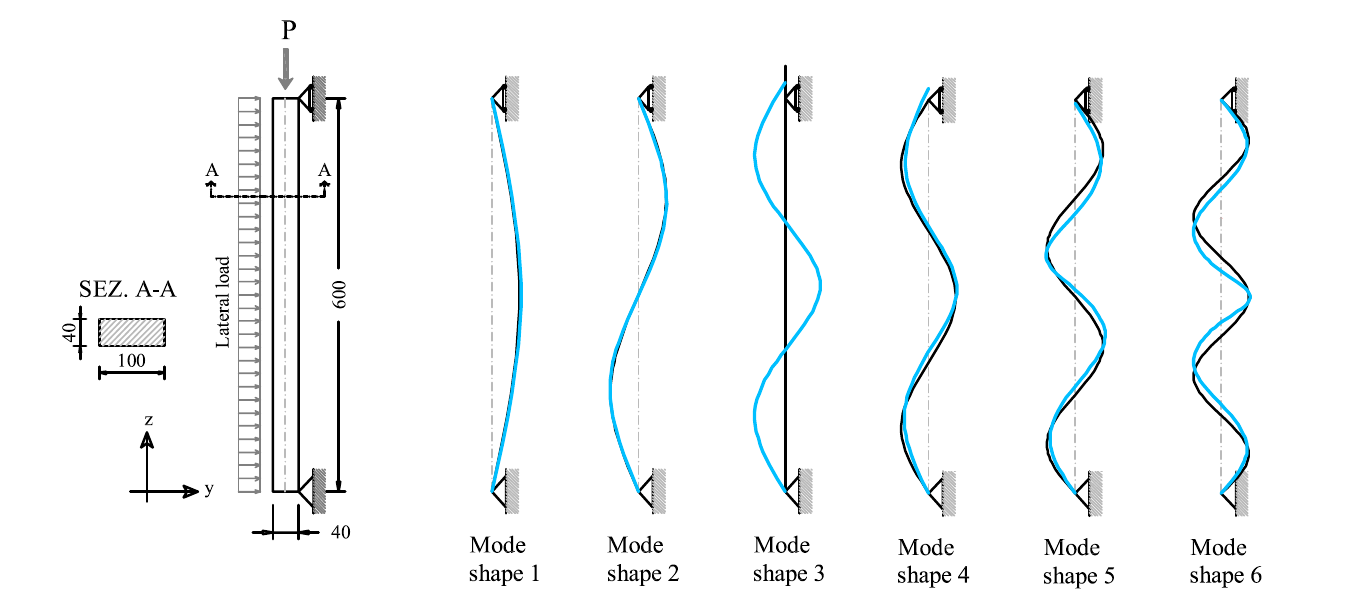}
\caption{The geometry of the beam (left) and
its first six mode shapes (right): linear elastic (black) and
masonry--like (cyan) case (lengths in cm). The masonry--like case is shown for the
last load increment.} \label{beam_modes}
\end{figure}

\begin{figure}\centering
\includegraphics[width=14cm]{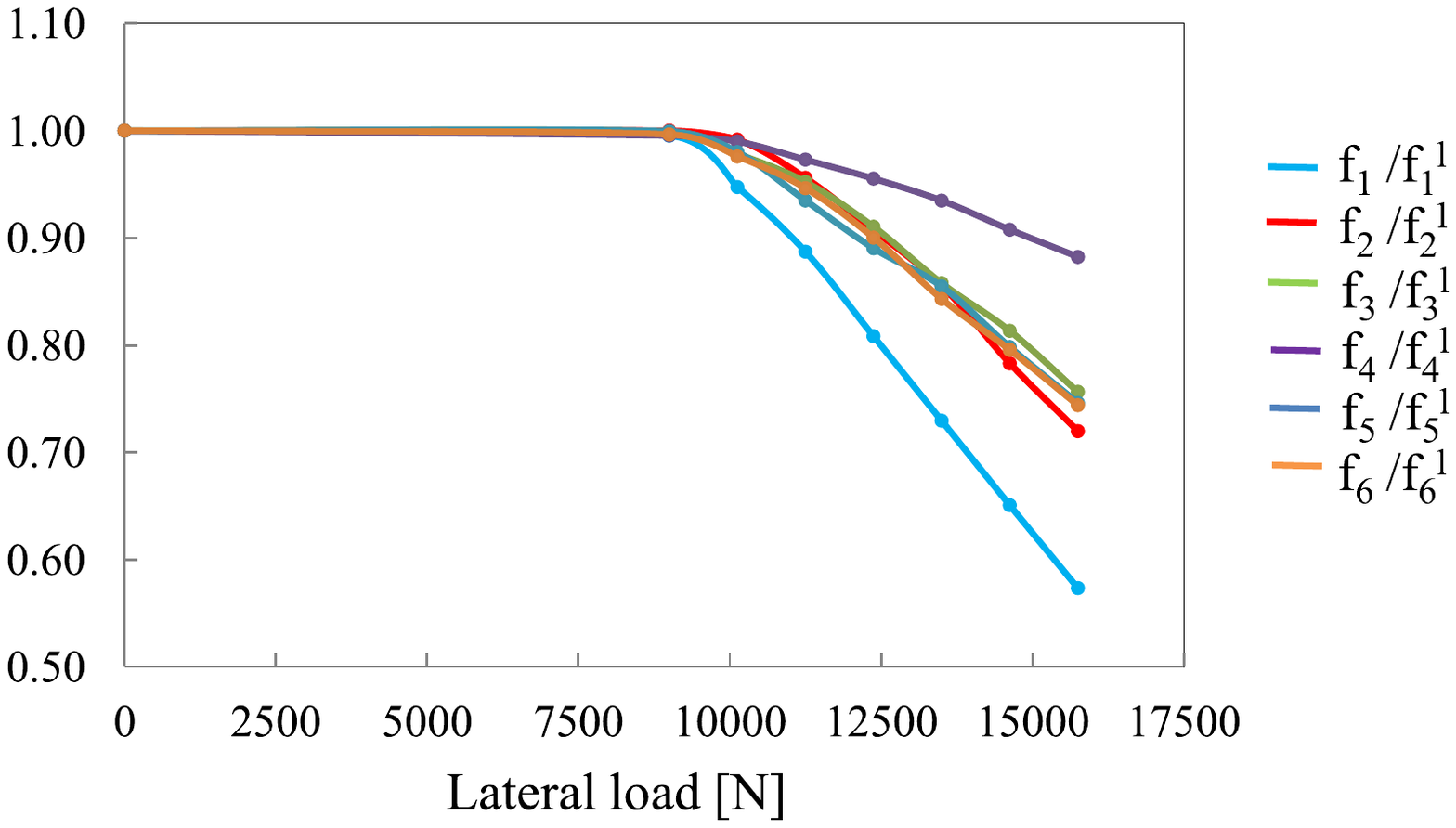}
\caption{The ratio $\text{f}_\text{i} / {\text{f}_\text{i}}^l$ for
the first six natural frequencies of the beam vs. the lateral
load.}\label{beam_frequencies}
\end{figure}

Let us consider the rectilinear beam illustrated in Figure
\ref{beam_modes}, with length $l=\unit{6}{\metre}$, rectangular
cross section of $\unit{0.4}{\metre}\times\unit{1}{\metre}$,
subjected to the concentrated vertical load $P=\unit{10^6}{\newton}$
and a uniform lateral load. The beam, simply supported at its ends
and forced to move in the $y-z$ plane, is discretized with $60$ beam
elements \cite{NOSA}. The goal of the finite--element analysis,
conducted with the NOSA--ITACA 1.2 code is to compare the natural
frequencies and mode shapes of the beam in the linear elastic case
with those in the presence of the damage induced by the increasing
lateral load. We conducted a preliminary modal analysis by assuming
the beam to be made of a linear elastic material with Young's
modulus $E=\unit{3\cdot10^9}{\pascal}$, Poisson's ratio $\nu=0.2$
and mass density $\rho=\unit{1800}{\kilogram\per\metre^3}$, for
which we calculated the corresponding natural frequencies
$\text{f}_\text{i}^l$ and mode shapes $\phi_\text{i}^l$. Then, by
following the procedure outlined in Section \ref{sec3}, we
incremented the lateral load and, at each increment, calculated the
frequencies $\text{f}_\text{i}$ and modes $\phi_\text{i}$. The value
of the lateral load applied to the beam was increased through seven
increments from $\unit{9\cdot10^3}{\newton}$ (which was able to
induce the first crack in the beam's midsection) to
$\unit{15.75\cdot10^3}{\newton}$. Table \ref{tab1} reports the value
of the lateral load and the corresponding natural frequencies. The
first row summarizes the results of the linear elastic modal
analysis. As outlined in Table \ref{tab1}, as the lateral load
increases, the beam's fundamental frequency decreases from the
linear elastic value of $\unit{6.47}{\hertz}$ to
$\unit{3.71}{\hertz}$, about one half its initial value. The other
frequencies fall by up to thirty per cent. The frequencies ratio
$\text{f}_\text{i}/\text{f}_\text{i}^l$ is reported in Figure
\ref{beam_frequencies}. The Figure clearly shows that when the
lateral load is applied, the fundamental frequency value falls
faster than the other frequency values. This is due to the chosen
lateral load distribution, which induces a deformation in the beam
similar to the first mode shape.  Table \ref{tab1} also reports the
values of the effective modal masses \cite{CP} calculated for the
first six mode shapes. The modes are also shown in Figure
\ref{beam_modes}, where the black line stands for the modes
$\phi_\text{i}^l$ calculated for the linear elastic case, while the
cyan line represents the modes $\phi_\text{i}$ calculated at the
last increment of the analysis. It is worth noting that the first,
second, fifth and sixth modes substantially maintain their shape,
while some changes occur between the third and fourth mode: their
frequency values are very similar, but in the linear elastic
analysis the third mode has an axial direction, while the fourth
mode has transverse direction. The nonlinear case shows that the
masses of the third mode pass from the axial to the transverse
direction, while for the fourth mode diverts from transverse to
axial \cite{NAY}. A small part of the total mass in the axial
direction migrates towards the higher modes and is not shown in the
Table.

\begin{figure}\centering
\includegraphics[width=12cm]{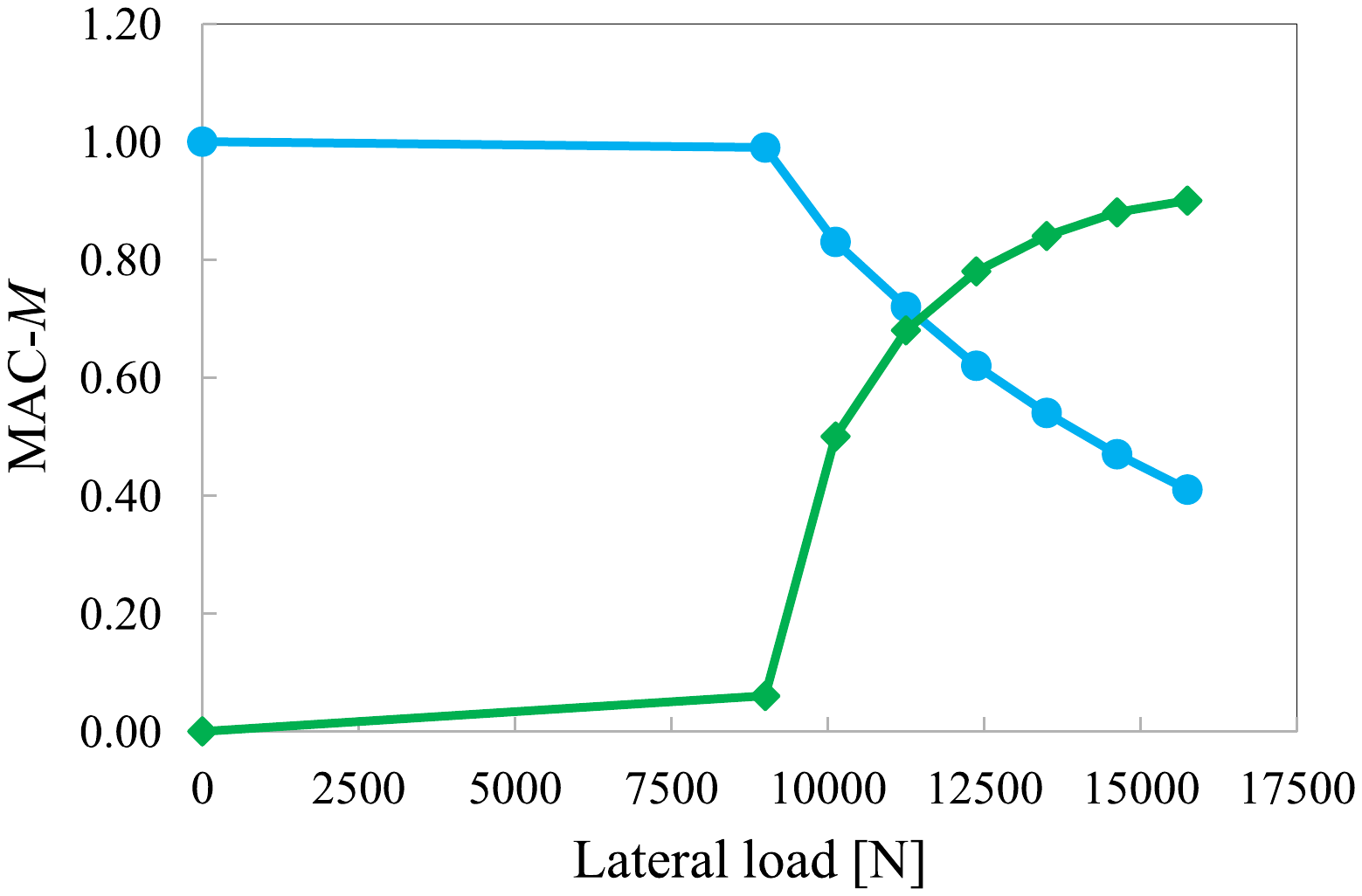}
\caption{The quantities MAC--$M (\phi_3^{l}, \phi_3)$ (blue line)
and MAC--$M (\phi_3^{l}, \phi_4)$ (green line)  vs. the lateral
load.} \label{MAC_beam}
\end{figure}

\begin{figure}\centering
\includegraphics[width=8cm]{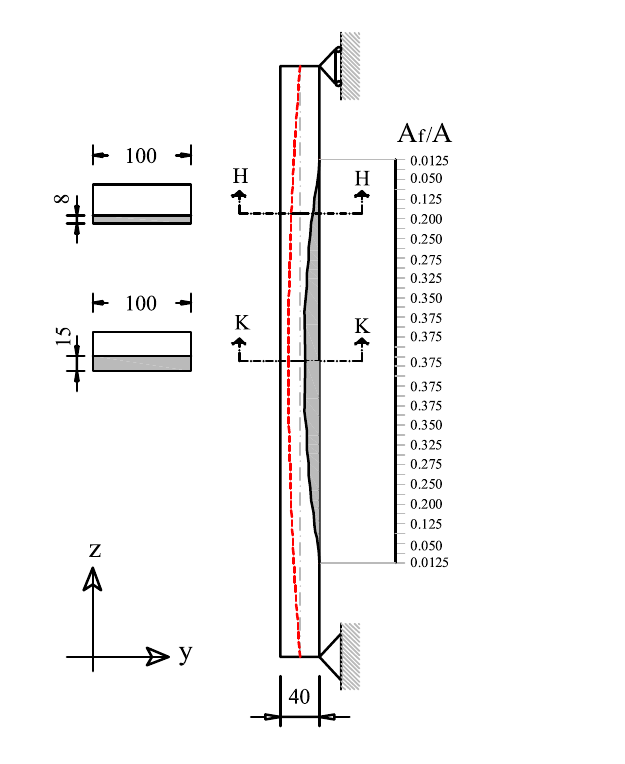}
\caption{The ratio between the cracked area $\text{A}_\text{f}$ and
total area  $\text{A}$ along the beam at the last load increment.
(lengths in cm).} \label{beam_cracked}
\end{figure}

In order to compare the mode shapes $\phi_\text{i}^l$ and
$\phi_\text{j}$, we introduce the quantity

\begin{equation}
\text{MAC--}M(\phi_\text{i}^l,\phi_\text{j})=\frac{|\phi_\text{i}^l\cdot
M\phi_\text{j}|}{\sqrt{\phi_\text{i}^l\cdot
M\phi_\text{i}}\sqrt{\phi_\text{j}^l\cdot M\phi_\text{j}}},
\end{equation}

\noindent which measures the correlation between the
$\text{i}-\text{th}$ linear elastic mode shape and the
$\text{j}-\text{th}$ mode shape of the damaged structure. More
precisely, if the MAC--$M$ value is close to unity, then the
vectors $\phi_\text{i}^l$ and $\phi_\text{j}$ are nearly parallel with
to respect to the scalar product in $\mathbb{R}^{n}$ induced by
matrix $M$. Table \ref{tab2} reports the quantities
MAC--$M(\phi_\text{i}^l,\phi_\text{j})$ calculated for i, j = 1, 2,
..., 6, with $\phi_\text{j}$ the damaged mode shape calculated at
the last load increment, corresponding to the lateral load
$\unit{15.75\cdot10^3}{\newton}$. As the lateral load increases, the
quantities MAC--$M(\phi_\text{3}^l,\phi_\text{3})$ and
MAC--$M(\phi_\text{3}^l,\phi_\text{4})$ drawn in Figure
\ref{MAC_beam} in blue and green lines, respectively, pass from $1$
to $0.4$ and from $0$ to $0.88$ (see Table \ref{tab2}), thus showing
that  mode shape $\phi_\text{3}$, initially coincident with
$\phi_\text{3}^l$, tends to become $M$-orthogonal to it and
$\phi_\text{4}$ initially $M$-orthogonal to $\phi_\text{3}^l$ tends
to become $M$-parallel to it. Finally, Figure \ref{beam_cracked}
shows the distributions of the fractured areas in the beam at the
last load increment, together with the ratio between the cracked
area $\text{A}_\text{f}$ and total area $\text{A}$ calculated in
every section of the beam.

\subsection{The masonry arch on piers} \label{sec4_2}

\begin{figure}
\includegraphics[width=14cm]{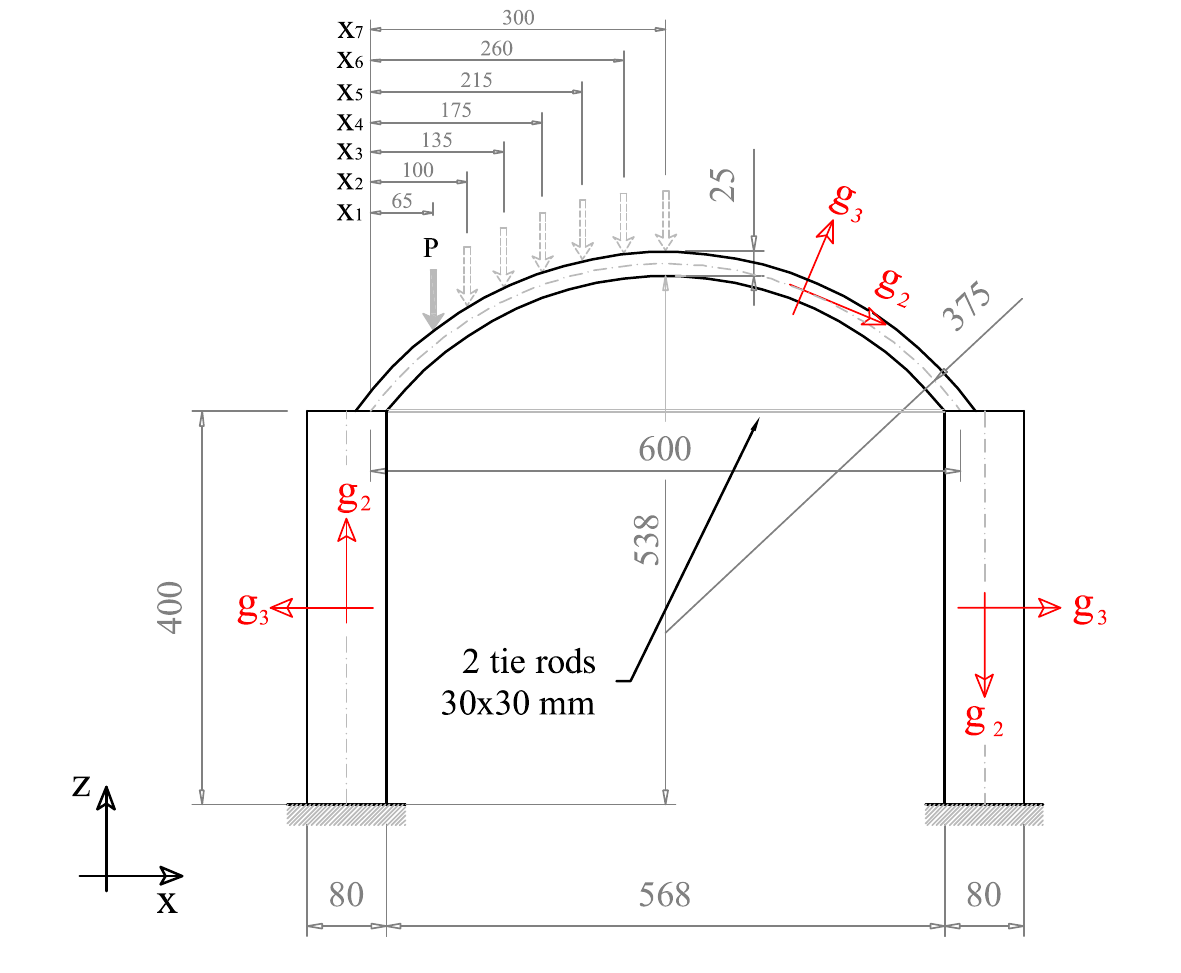}
\caption{Geometry of the structure made of up the arch on piers and
the different positions  along the arch span at which the
concentrated load P is applied (length in cm). The local axes of the
finite--elements are drawn in red.} \label{arch_geometry}
\end{figure}

Let us consider the masonry arch on piers  shown in  Figure
\ref{arch_geometry}. The arch span is $\unit{6}{\metre}$ and its
cross section  measures $\unit{0.25}{\metre}\times\unit{1}{\metre}$.
The arch has a circular shape, with a mean radius of
$\unit{3.75}{\metre}$, and rests on two lateral piers with
rectangular cross section of
$\unit{0.8}{\metre}\times\unit{1}{\metre}$ and height
$\unit{4}{\metre}$. The structure is reinforced by means of two
steel tie rods, with rectangular cross sections of
$\unit{0.03}{\metre}\times\unit{0.03}{\metre}$, fixed at the
pier--arch nodes. An offset of $\unit{0.24}{\metre}$ has been
considered between the axis of the arch and those of the piers. The
structure, modelled by means of $1500$ thick shell elements,
\cite{NOSA}, \cite{Voltone}, is clamped at the piers' base and
forced to move in the $x-z$ plane. Figure \ref{arch_geometry} also
shows
 the elements' local axes in red. Beam elements \cite{NOSA} have
been used to model the tie rods.

Just as for the beam described in the previous subsection, we
conducted a preliminary modal analysis by assuming the masonry
structure to be made of a linear elastic material with Young's
modulus $E=\unit{3\cdot10^9}{\pascal}$, Poisson's ratio $\nu=0.2$
and mass density $\rho=\unit{1800}{\kilogram\per\metre^3}$, while
for the tie rods we have assumed
$E=\unit{2.1\cdot10^{11}}{\pascal}$, Poisson's ratio $\nu=0.3$ and
mass density $\rho=\unit{7850}{\kilogram\per\metre^3}$. The modal
analysis allowed us to calculate the structure's natural frequencies
$\text{f}_\text{i}^l$ and mode shapes $\phi_\text{i}^l$. The loads
have been applied incrementally, first the self--weight of the
structure alone, and then a concentrated vertical load P, whose
value was
 increased through four increments from $\unit{150.61\cdot10^3}{\newton}$ to
$\unit{154.61\cdot10^3}{\newton}$. Seven
analyses were performed, each time moving the load P to different
positions along the arch span (see Figure \ref{arch_geometry}). The natural frequencies
$\text{f}_\text{i}$ and mode shapes $\phi_\text{i}$ of the damaged
structure have thus been calculated for
each analysis at each load increment.

\begin{figure}\centering
\includegraphics[width=14cm]{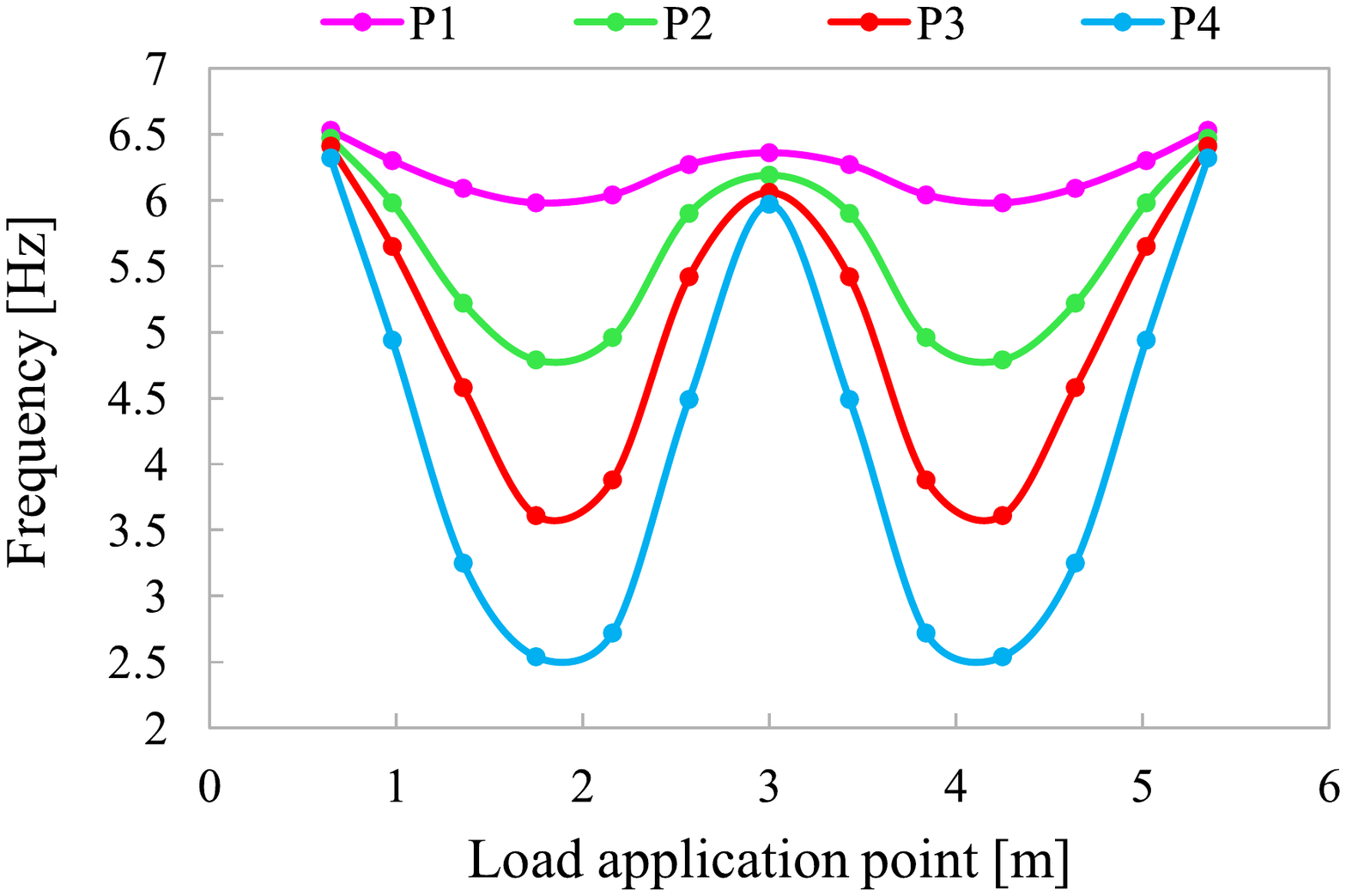}
\caption{The structure's frequency $\text{f}_\text{1}$ vs. the
position of the concentrated load P for different load increments.}
\label{arch_f1}
\end{figure}

\begin{figure}\centering
\includegraphics[width=14cm]{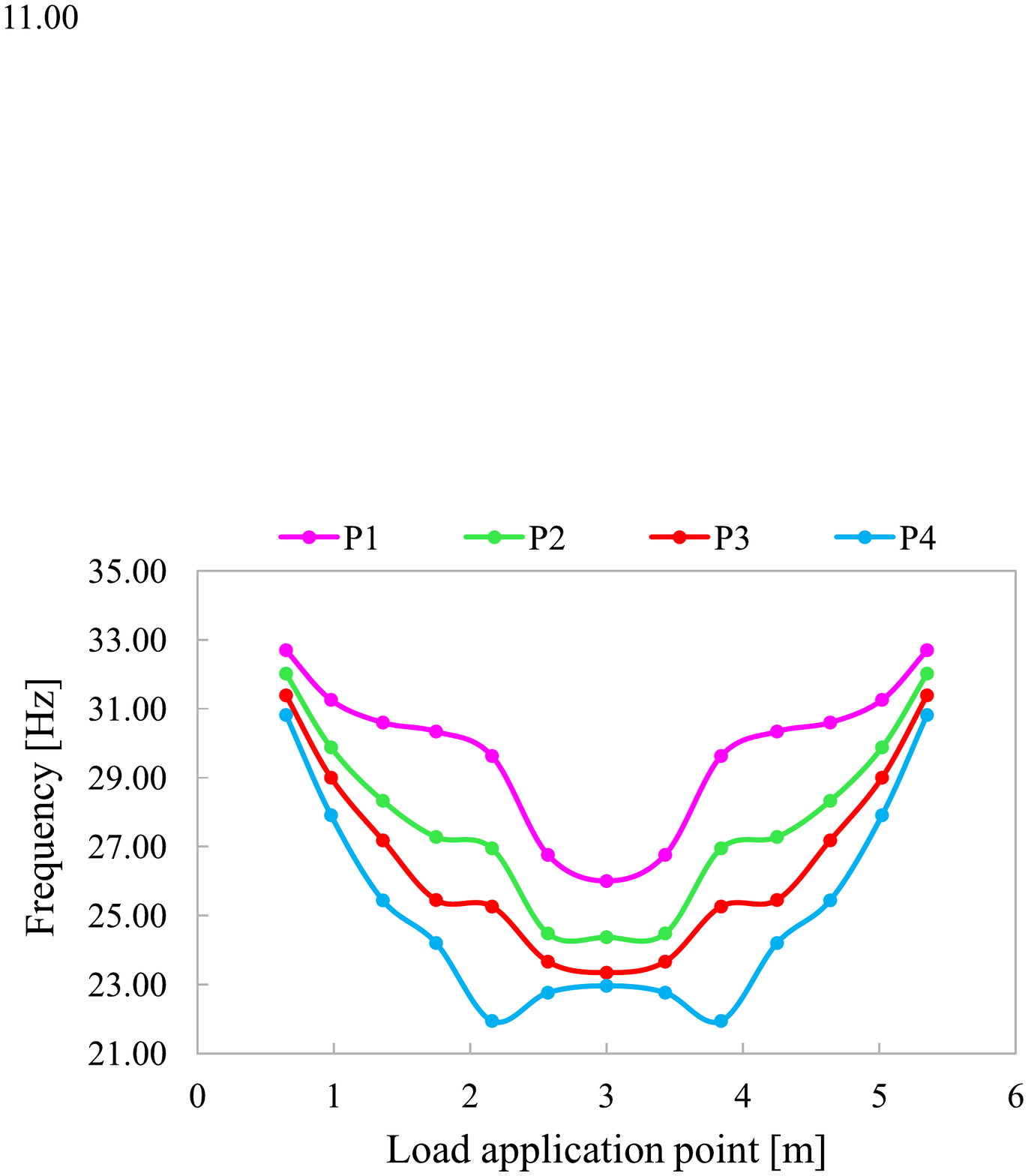}
\caption{The structure's frequency $\text{f}_\text{4}$ vs. the
position of the concentrated load P for different load increments.}
\label{arch_f4}
\end{figure}

\begin{figure}\centering
\includegraphics[width=14cm]{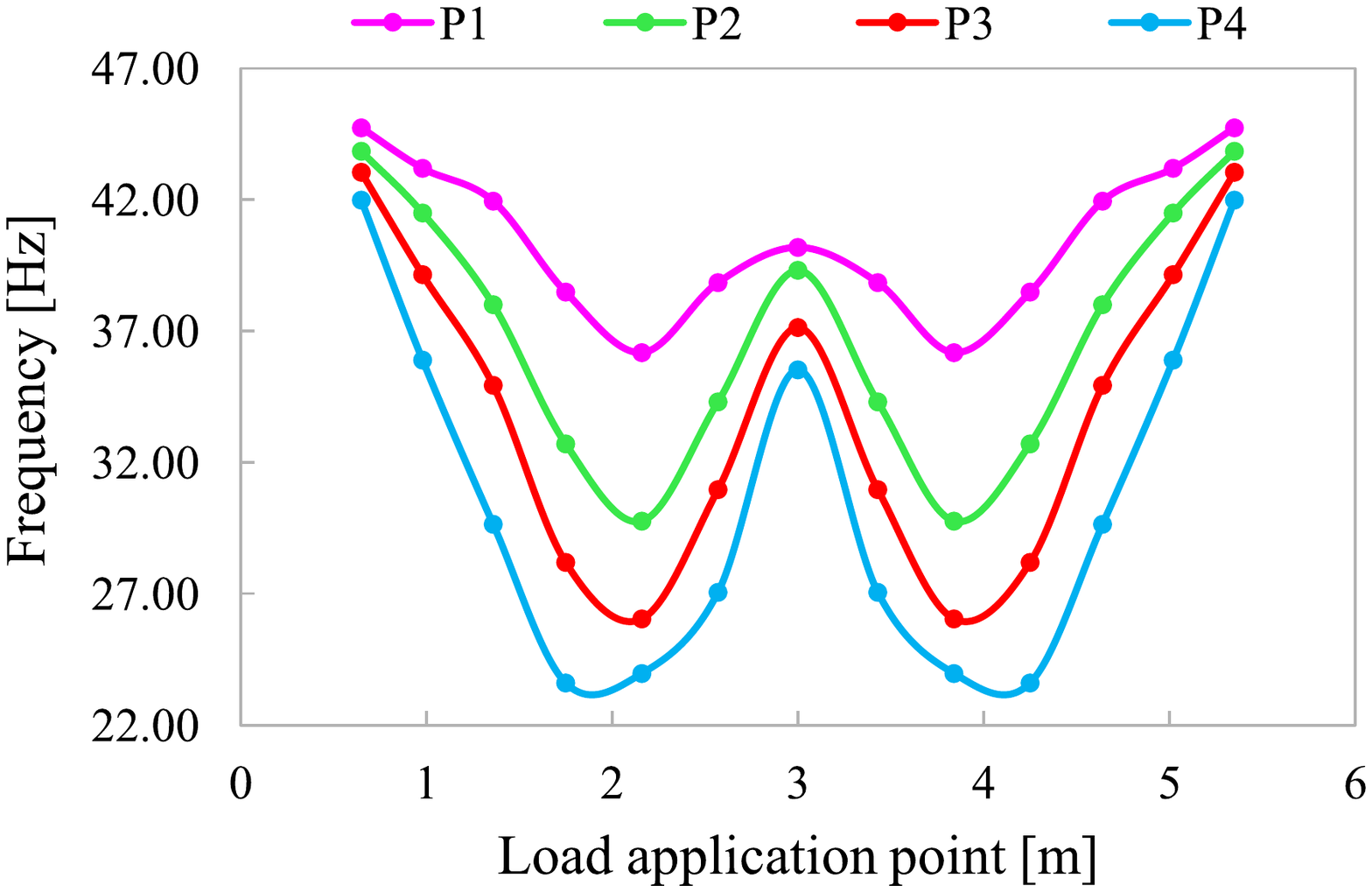}
\caption{The structure's frequency $\text{f}_\text{5}$ vs. the
position of the concentrated load P for different load increments.}
\label{arch_f5}
\end{figure}

Table \ref{tab3} shows the structure's first five natural
frequencies vs. load increment for the concentrated load P in
$X_4$, at about one fourth of the arch span. As the load P increases,
the fundamental frequency value of the structure decreases from
$\unit{6.54}{\hertz}$ to $\unit{2.54}{\hertz}$, the others fall by up
to fifty percent of their linear elastic value. This behaviour is also evident
 in Table \ref{tab5}, for P in the middle of the arch. Figures
\ref{arch_f1} to \ref{arch_f5} show the first five frequencies
of the structure vs. the load application point along the arch span,
for the four load P values used during the analyses
($\text{P}_1=\unit{150605}{\newton}$,
$\text{P}_2=\unit{152605}{\newton}$,
$\text{P}_3=\unit{153605}{\newton}$,
$\text{P}_4=\unit{154605}{\newton}$). These Figures show that all
the structure's frequencies reach their minimum values when
load P is at the arch's quarter points (between positions $X_4$ and
$X_5$). The minimum decrease with respect to the linear elastic case
is exhibited by frequencies $\text{f}_\text{1}, \,
\text{f}_\text{2},\,\text{f}_\text{5}$ for P at $X_7$, while for
frequencies $\text{f}_\text{3}, \, \text{f}_\text{4}$ it appears
between positions $X_5$ and $X_6$. For a given load position $X_i$,
all frequency values decrease their values as the load increases.

\begin{figure}\centering
\includegraphics[width=16cm]{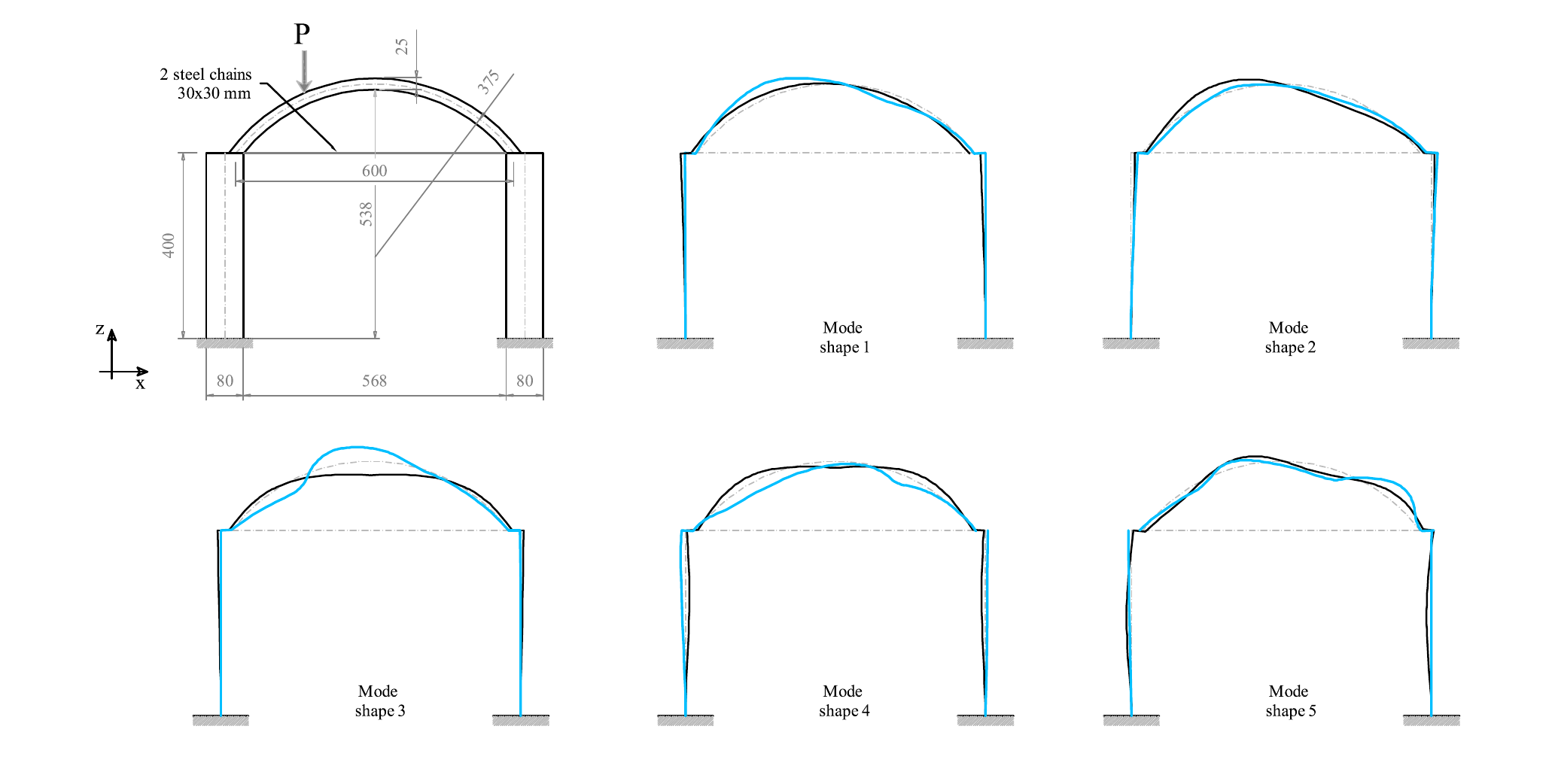}
\caption{The  first five mode shapes of the structure for load P at
position $X_4$: linear elastic (black) and masonry--like (cyan)
case. The masonry--like case is shown for the last load increment
(length in cm).} \label{arch_x4_modes}
\end{figure}

\begin{figure}\centering
\includegraphics[width=16cm]{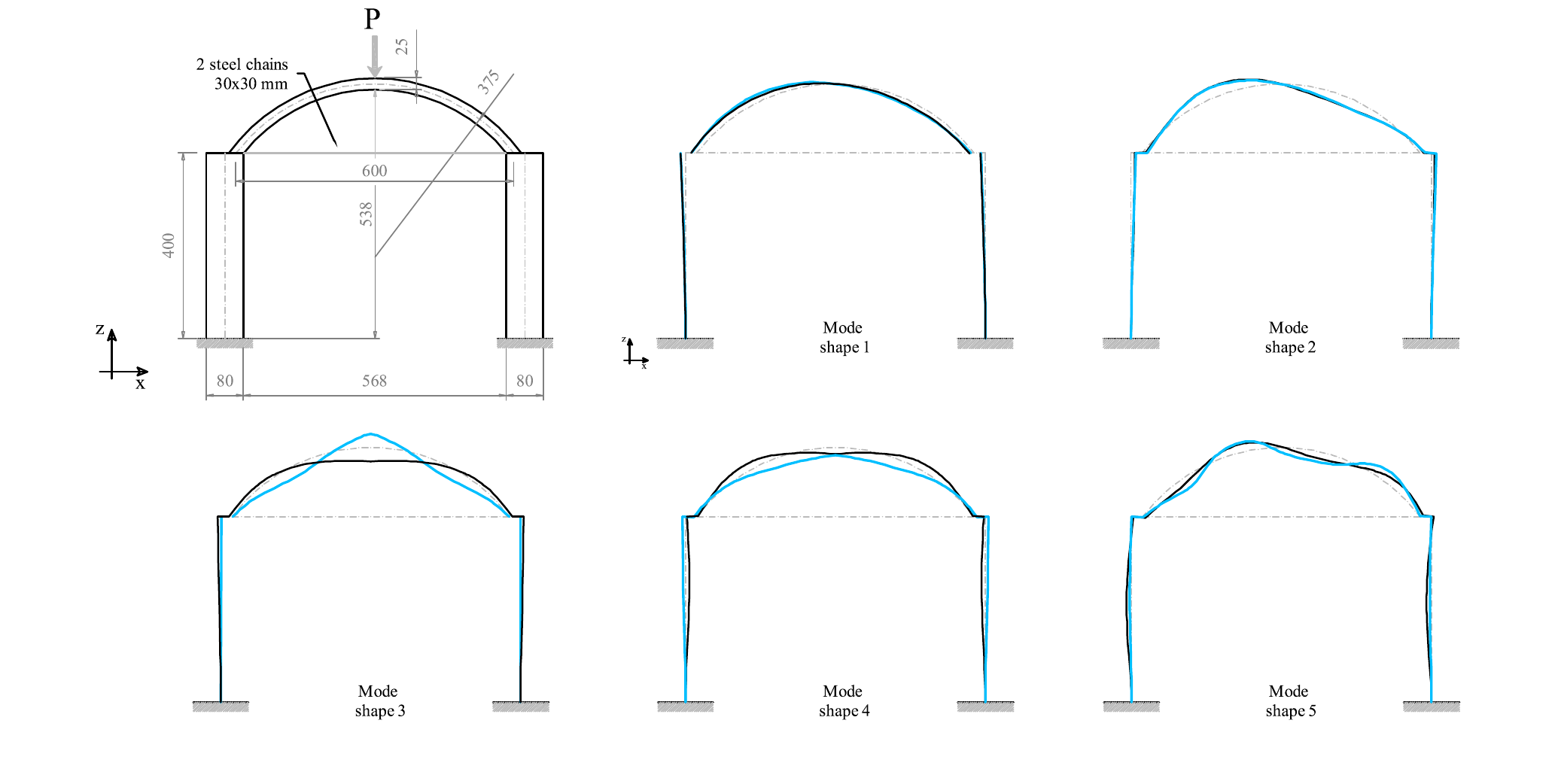}
\caption{The  first five mode shapes of the structure for load P at
position $X_7$: linear elastic (black) and masonry--like (cyan)
case. The masonry--like case is shown for the last load increment.}
\label{arch_x7_modes}
\end{figure}

The structure's mode shapes  are shown in Figures
\ref{arch_x4_modes} and \ref{arch_x7_modes} for P at $X_4$ and
$X_7$, respectively. The corresponding effective modal masses are
shown vs. load increments in Table \ref{tab3} and \ref{tab5},
respectively. The masses excited by the first five modes in the
linear elastic case \cite{karnovsky} are about $80\%$ of the total
mass in the $x$ direction and $10\%$ in the $z$ direction. The $x$
direction, prevalent in modes $1$, $2$ and $5$, is related to
oscillations involving movement of the entire structure, while the $z$
direction, prevalent in modes $3$ and $4$, relates to modes mainly
involving local oscillations of the arch and piers. The values
of the total excited masses remain stable during the nonlinear
analyses, but their distribution among the modes tends to change: in
particular, the masses excited along $x$ tend to migrate from the
first mode shape to the second, while the masses along $z$ pass from
mode shape $3$ to mode shape $4$. This phenomenon is particularly
evident when load P is at $X_4$. Accordingly, the MAC--$M$
matrix plotted in Table \ref{tab4} for P at $X_4$ exhibits low
values of the diagonal terms, while the off--diagonal terms reveal
a high correlation at the end of the analysis between the first and
second and third and fourth modes. Table \ref{tab6}, for P at
$X_7$, shows that the first, second and fifth modes
substantially maintain their shape, while the third and fourth do
not.

For each element we now introduce the Frobenius norm of the difference
between the elemental stiffness matrices $\tilde{K}_e$ and $K_e$,
whose assemblages form matrices $\tilde{K}$ and $K$,

\begin{equation}
d_e=\Vert \tilde{K}_e-K_e\Vert.
\end{equation}

\begin{figure}\centering
\includegraphics[width=18cm]{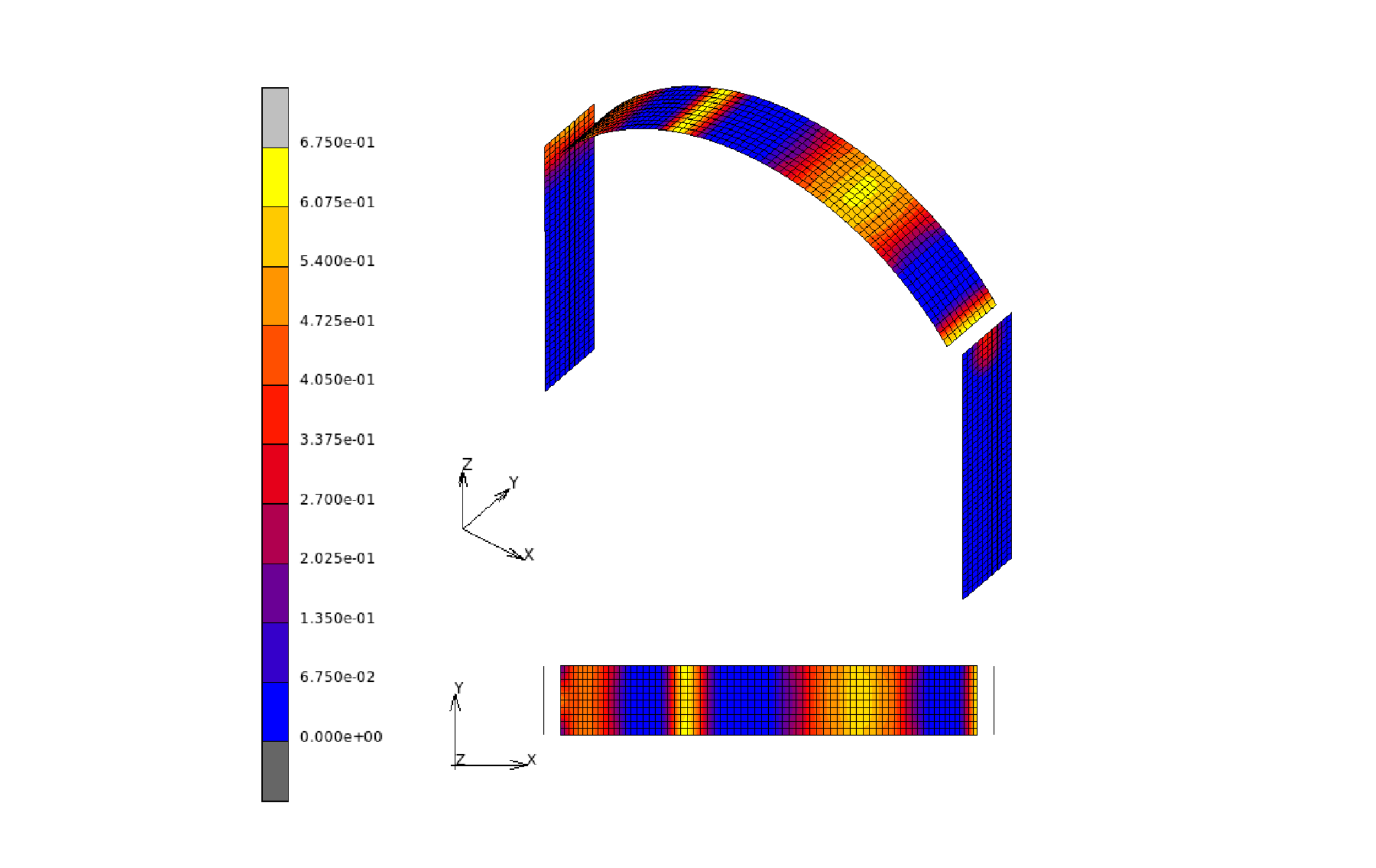}
\caption{Distribution of $d_e$, $e=1,..,1500$ in the structure for
the load P at position $X_4$. The result refers to the last
load increment. } \label{arch_D_x4}
\end{figure}

\begin{figure}\centering
\includegraphics[width=16cm]{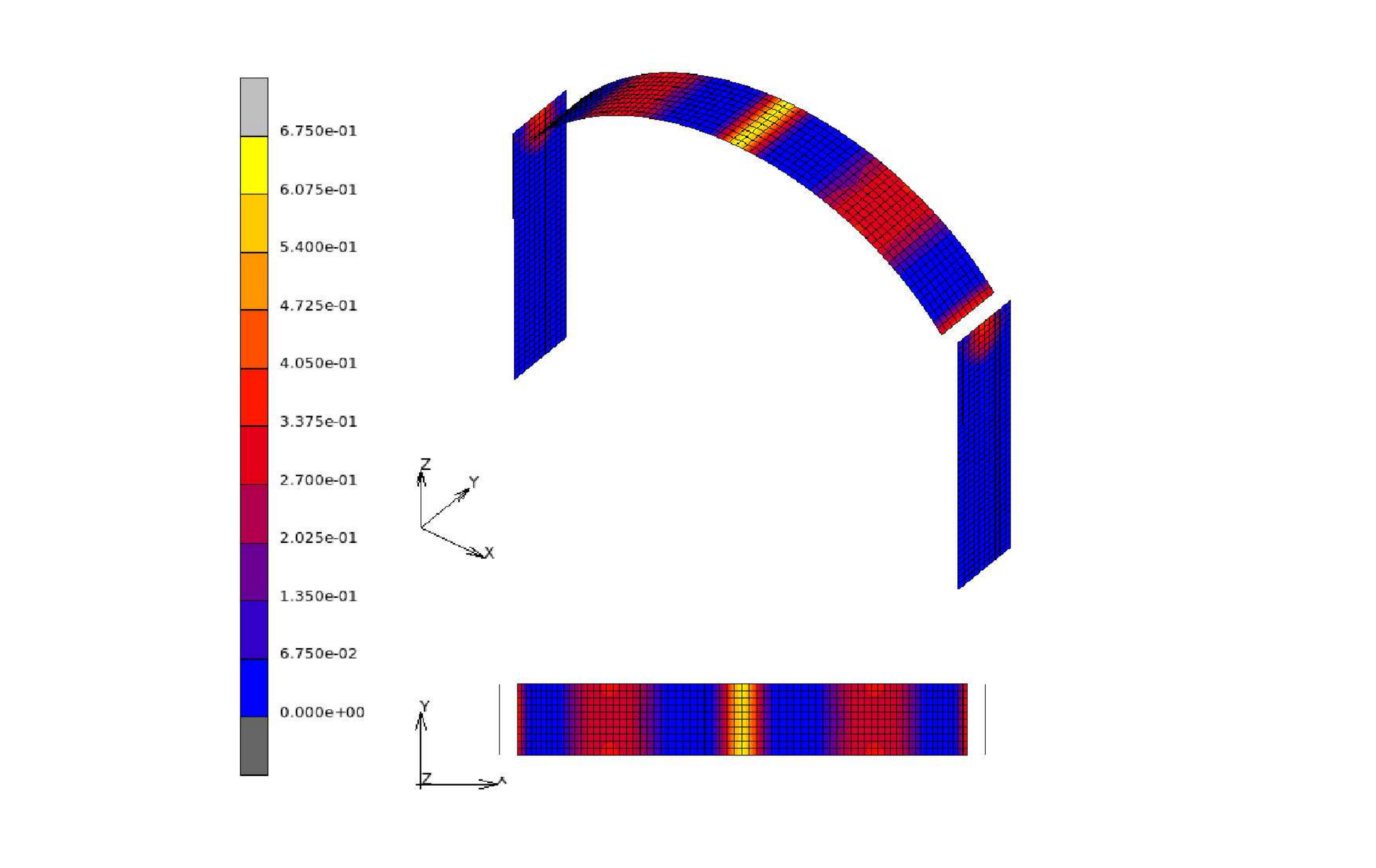}
\caption{Distribution of $d_e$, $e=1,..,1500$ in the structure for
the load P at position $X_7$. The result refers to the last
load increment.} \label{arch_D_x7}
\end{figure}

\begin{figure}\centering
\includegraphics[width=16cm]{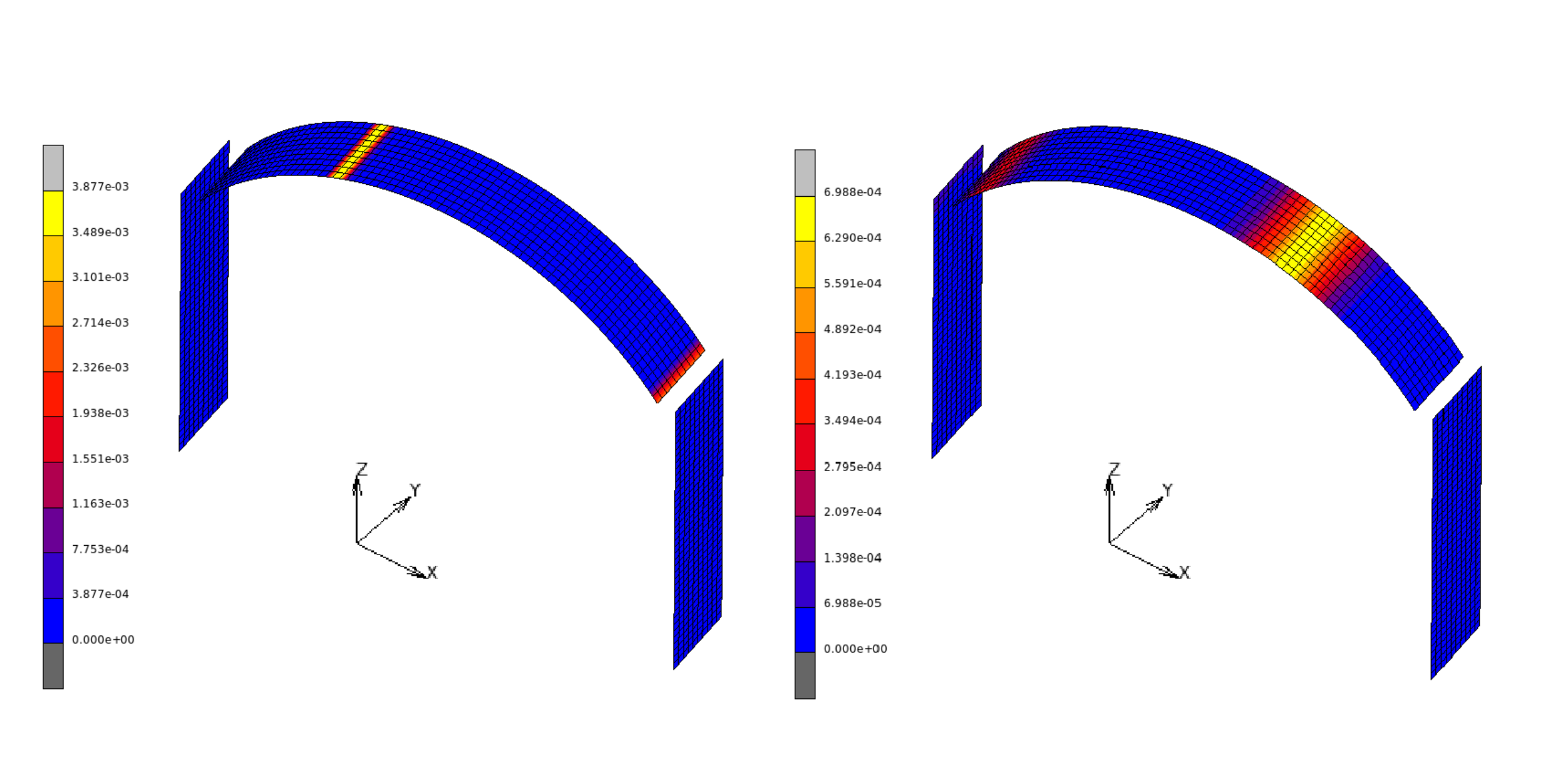}
\caption{Fracture strain $E_{22}^f$ at the intrados (left) and the
extrados (right) of the structure for load P at position $X_4$.}
\label{ef_x4}
\end{figure}

\begin{figure}\centering
\includegraphics[width=16cm]{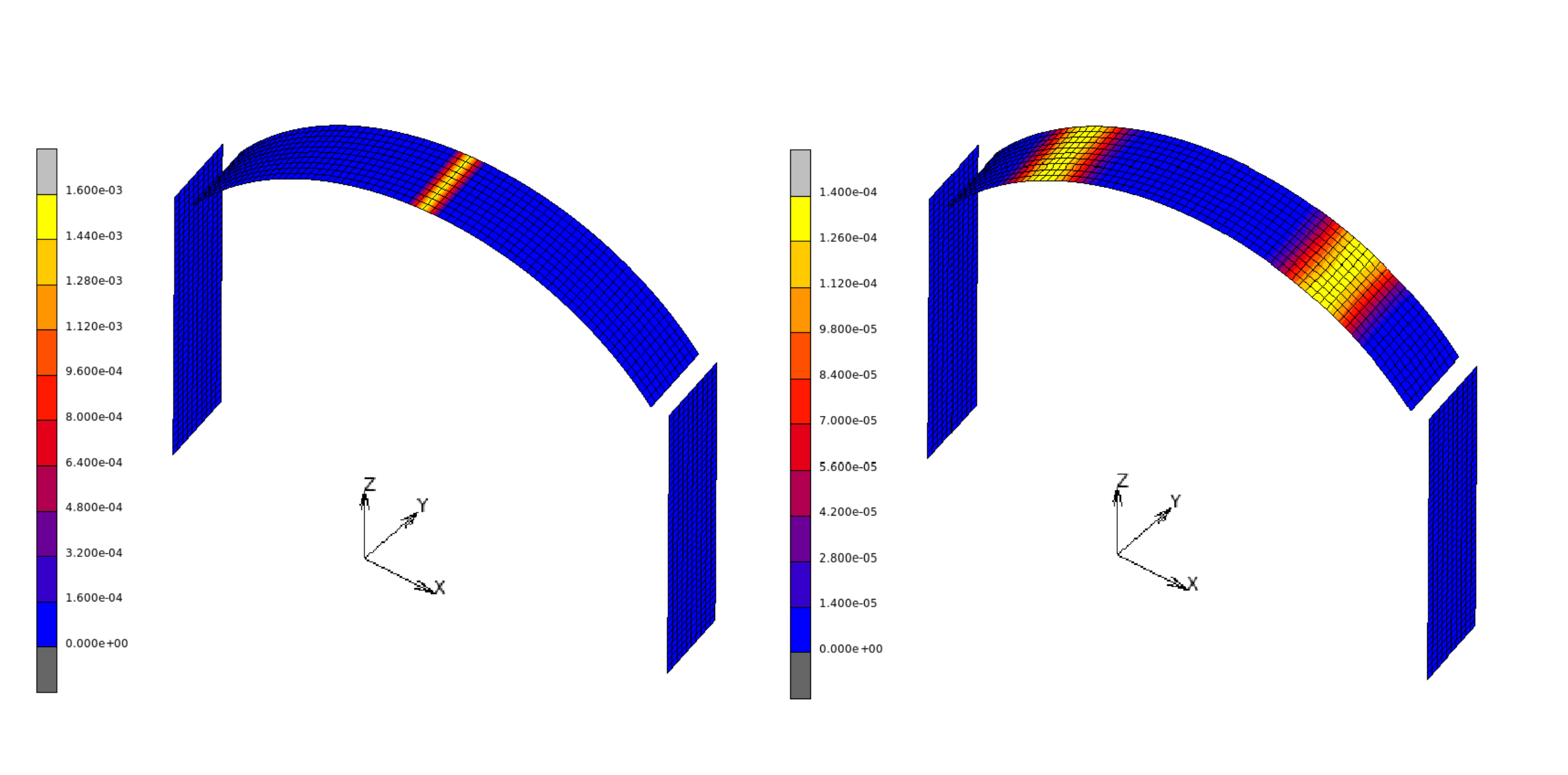}
\caption{Fracture strain $E_{22}^f$ at the intrados (left) and the
extrados (right) of the structure for load P at position $X_7$.}
\label{ef_x7}
\end{figure}

Figures \ref{arch_D_x4} and \ref{arch_D_x7} show the behaviour of
$d_e$, $e=1,..,1500$ and highlight the elements in which
the distance between the damaged and the linear elastic stiffness
matrices attains the highest values. These elements substantially
coincide with those characterized by the highest values of the
fracture strain, as shown in Figures \ref{ef_x4} and \ref{ef_x7},
where the strains $E_{22}^f=\mathbf{g}_2\cdot \mathbf{E}^f
\mathbf{g}_2$ at the intrados and extrados of the structure are
shown for P at $X_4$ and $X_7$, respectively.

\pagebreak


\begin{figure}\centering
\includegraphics[width=14cm]{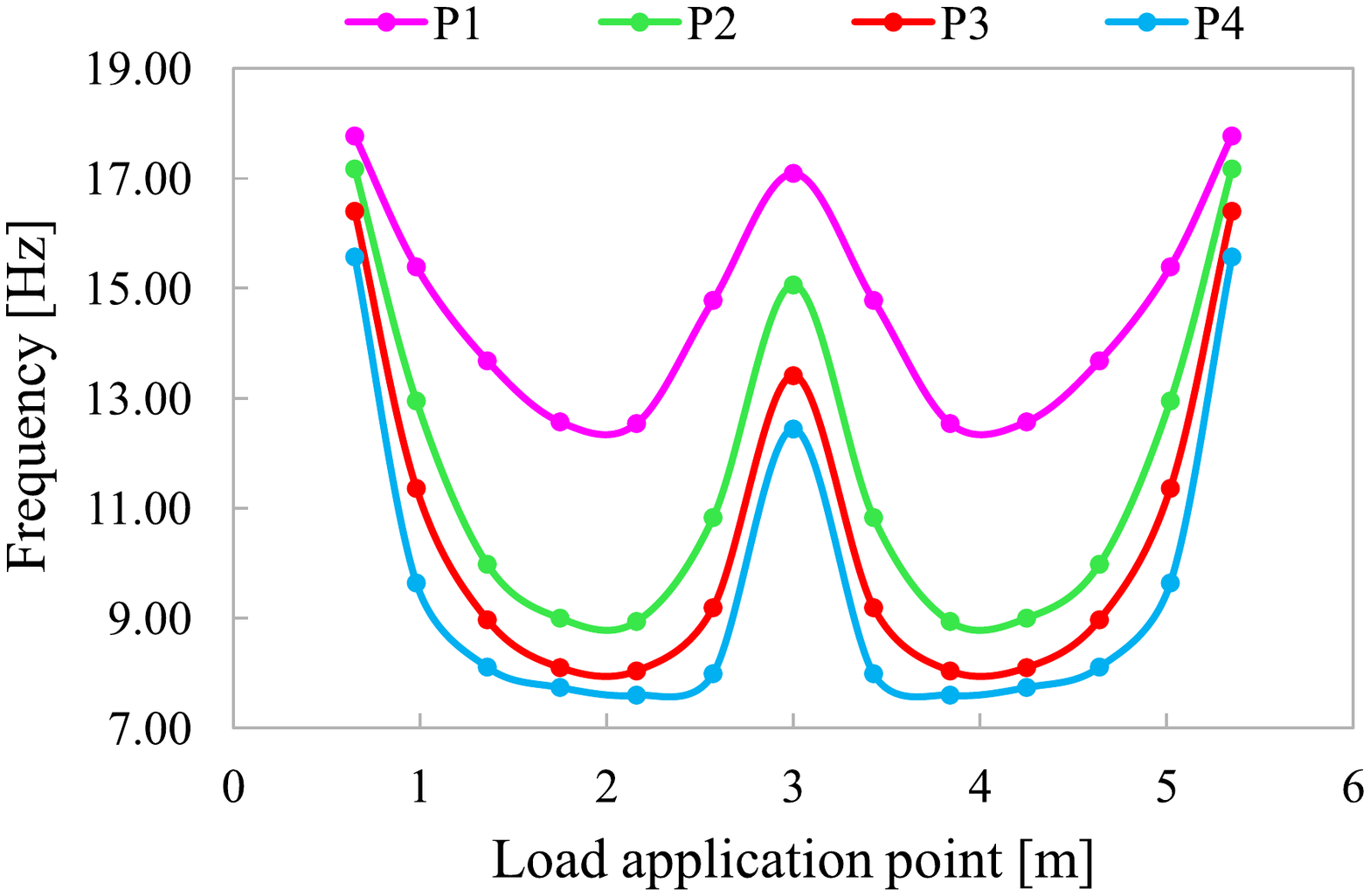}
\caption{The structure's frequency $\text{f}_\text{2}$ vs. the
position of the concentrated load P for different load increments.}
\label{arch_f3}
\end{figure}

\begin{figure}\centering
\includegraphics[width=14cm]{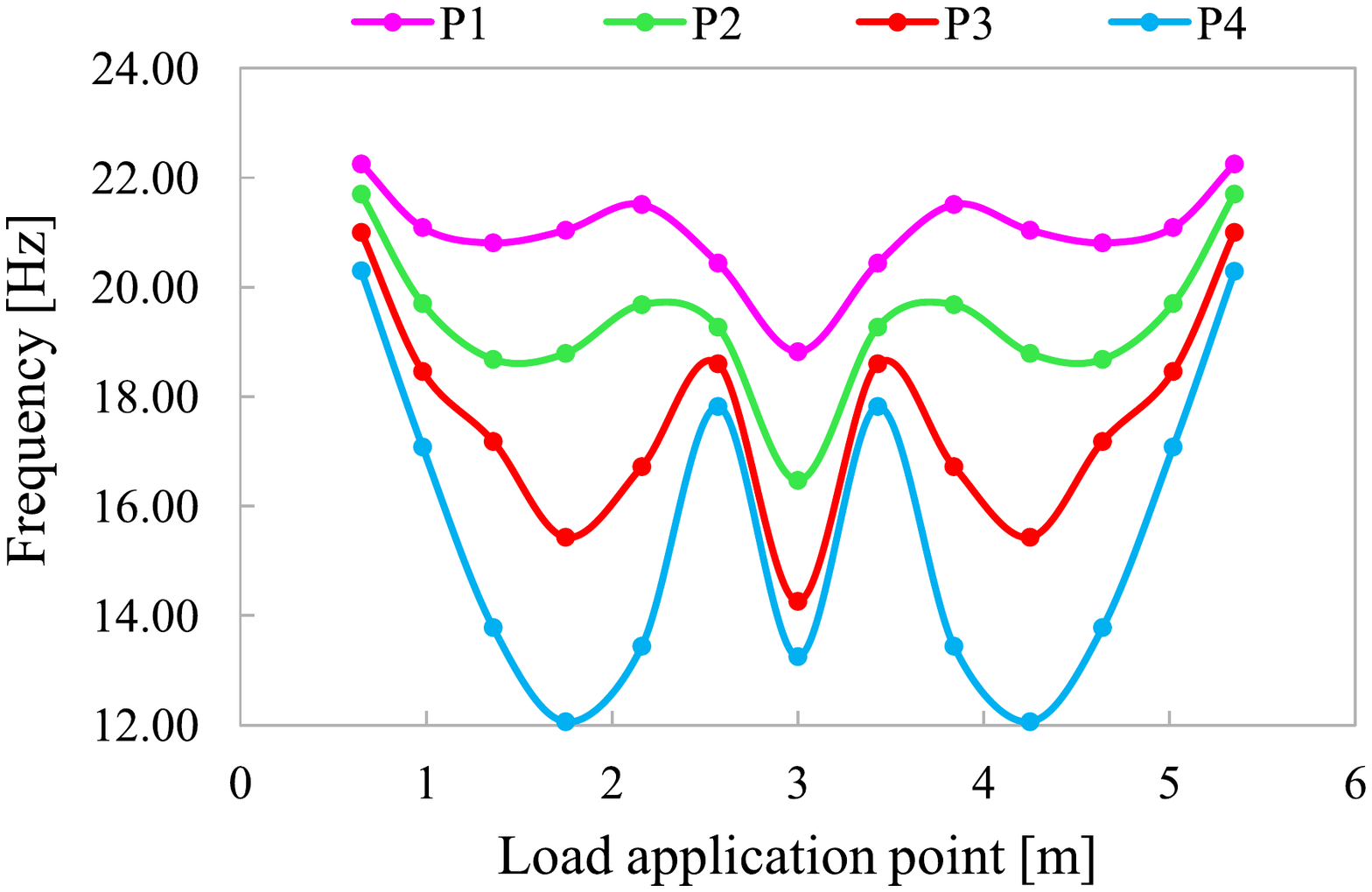}
\caption{The structure's frequency $\text{f}_\text{3}$ vs. the
position of the concentrated load P for different load increments.}
\label{arch_f3}
\end{figure}

\subsection{The bell tower of the Church of San Frediano in Lucca} \label{sec4_3}


\begin{figure}\centering
\includegraphics{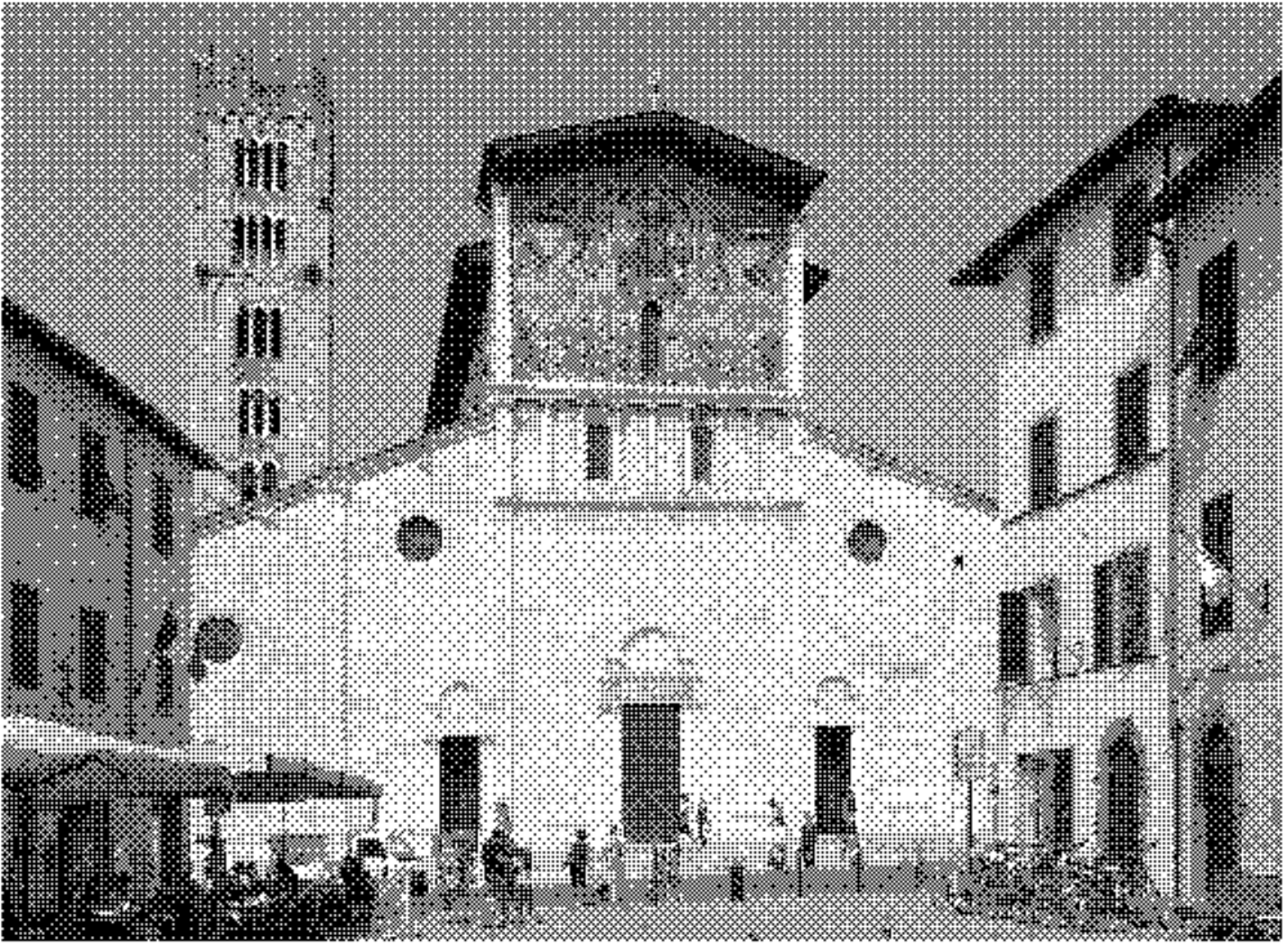}
\caption{The Basilica of San Frediano in Lucca (Italy). On the left,
the San Frediano bell tower.} \label{sanfrediano}
\end{figure}

\begin{figure}\centering
\includegraphics[width=15cm]{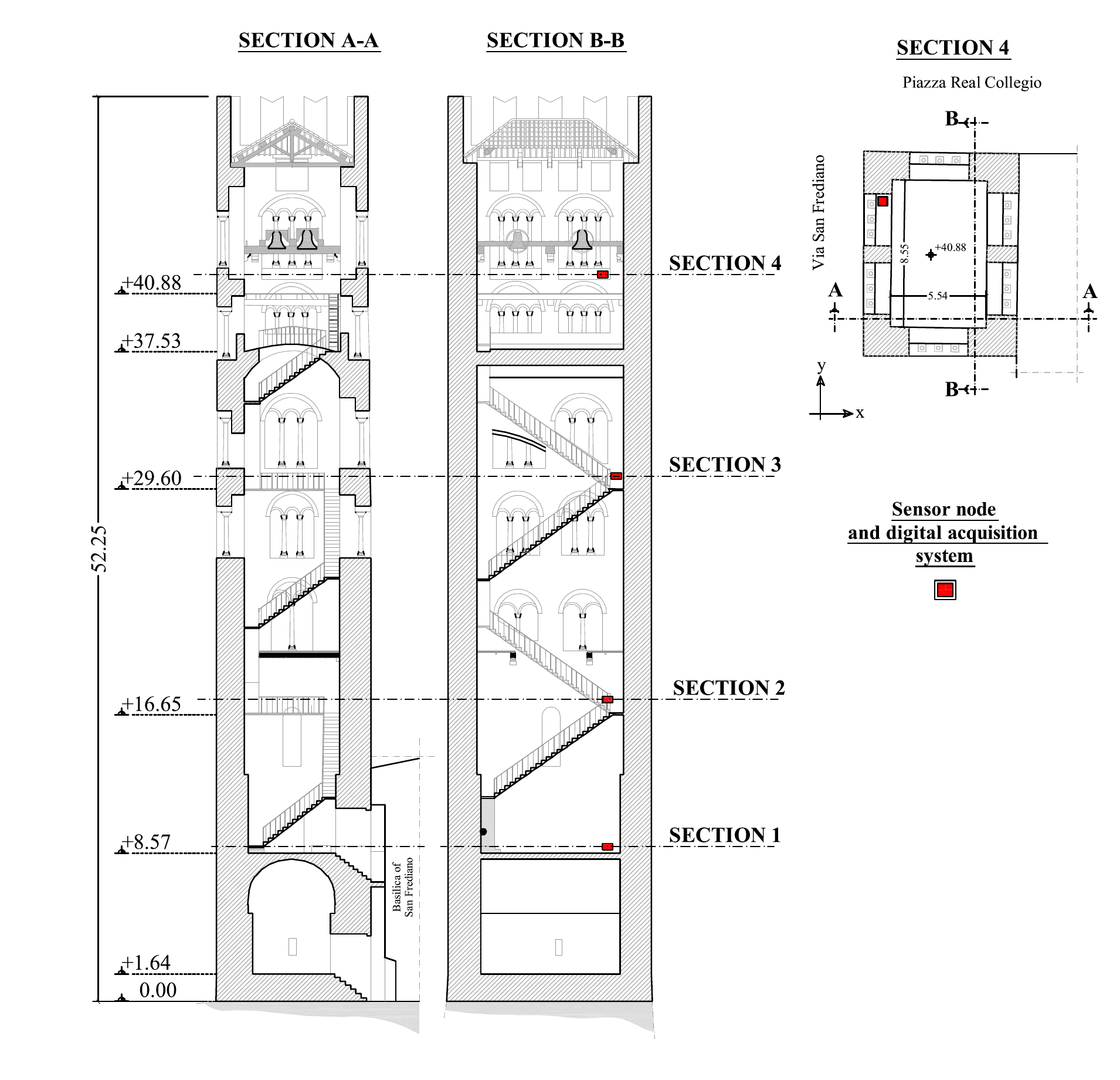}
\caption{Sections of the San Frediano bell tower and sensors'
arrangement.} \label{sanfrediano_sezioni}
\end{figure}

The Basilica of San Frediano (Figure \ref{sanfrediano}), dating back
to the 11th century, is one of the most fascinating monuments in
Lucca, much of its fascination being due to the marvellous mosaics
that adorn its façade. In June 2015 the church's 52 m high bell
tower was fitted with four high--sensitivity triaxial
seismometric stations, made available by the Arezzo Earthquake Observatory (Osservatorio Sismologico di Arezzo) and left active on the tower for three days
(Figure \ref{sanfrediano_sezioni}). Data from the
instruments, analyzed via OMA techniques \cite{BRINCKER}, allowed us
to determine the tower's first five natural frequencies
$f_\text{i}^\text{exp}$ and corresponding damping
ratios \cite{sanfrediano}. Information on the mode shapes was also
extracted, though the arrangement of the sensors, aligned along
the vertical, did not allow for reliable identification of the
torsional modes.

In the following the NOSA--ITACA code, together with model updating techniques,
is employed in order to fit the experimental results in the linear
elastic and the masonry--like (nonlinear) case. The
finite--element model of the tower consists of $18495$ thick shell
elements \cite{Voltone}. The four steel tie rods fitted to the masonry
vault under the bell chamber (between section $3$ and section $4$ in
Figure \ref{sanfrediano_sezioni}) are modelled with beam elements
\cite{NOSA}, which have also been used to model the wooden trusses and rafters
constituting the pavillion roof covering the tower. The
mechanical properties of the tie rods and wooden elements are,
respectively, $E = 2.1 \cdot 10^5$ MPa, $\nu = 0.3$, $\rho =
\unit{7850}{\kilogram\per\metre^3}$ and $E = 9500$ MPa, $\nu = 0.4$,
$\rho = \unit{800}{\kilogram\per\metre^3}$. With regard to the
materials constituting the masonry tower, no experimental
information is available to date. Visual inspection reveals that the
external layers of the walls are made up of regular stone blocks at
the base, while quite homogeneous brick masonry forms the
upper part, except for the central part of the walls, where the
masonry between the windows is made up of stone blocks. A first
attempt to fit the experimental results through model updating
procedures, conducted in \cite{sanfrediano}, showed that good
results can be achieved by assuming homogeneous values for the
mechanical properties of all the materials making up the tower.
Herein we proceed in the same way by assuming Young's
modulus $E$ and the mass density $\rho$ of masonry as parameters to
be updated. In particular, for different values of $E$ and $\rho$
satisfying

\begin{eqnarray}
\unit{3\cdot 10^9\,}{\pascal}\le E \le \unit{7\cdot 10^9\,}{\pascal},\nonumber\\
\unit{1800}{\kilogram\per\metre^3}\le \rho \le
\unit{2200}{\kilogram\per\metre^3}, \label{interval}
\end{eqnarray}

\noindent the first five natural frequencies of the tower are
calculated for the linear elastic and  nonlinear case, after having
applied the structure's self--weight alone. The optimal values of
$E$ and $\rho$,  are determined by minimizing the functions

\begin{equation}
e^l(E, \rho)=\sum_{i=1}^5
\left(f_\text{i}^l(E,\rho)-f_\text{i}^\text{exp}\right)^2,
\label{epslin}\end{equation}

\begin{equation}
e(E, \rho)=\sum_{i=1}^5
\left(f_\text{i}(E,\rho)-f_\text{i}^\text{exp}\right)^2,
\label{epsnonlin}\end{equation}

\noindent respectively in the linear and nonlinear cases, for $E$
and $\rho$ satisfying \eqref{interval}.

\begin{figure}\centering
\includegraphics[width=15cm]{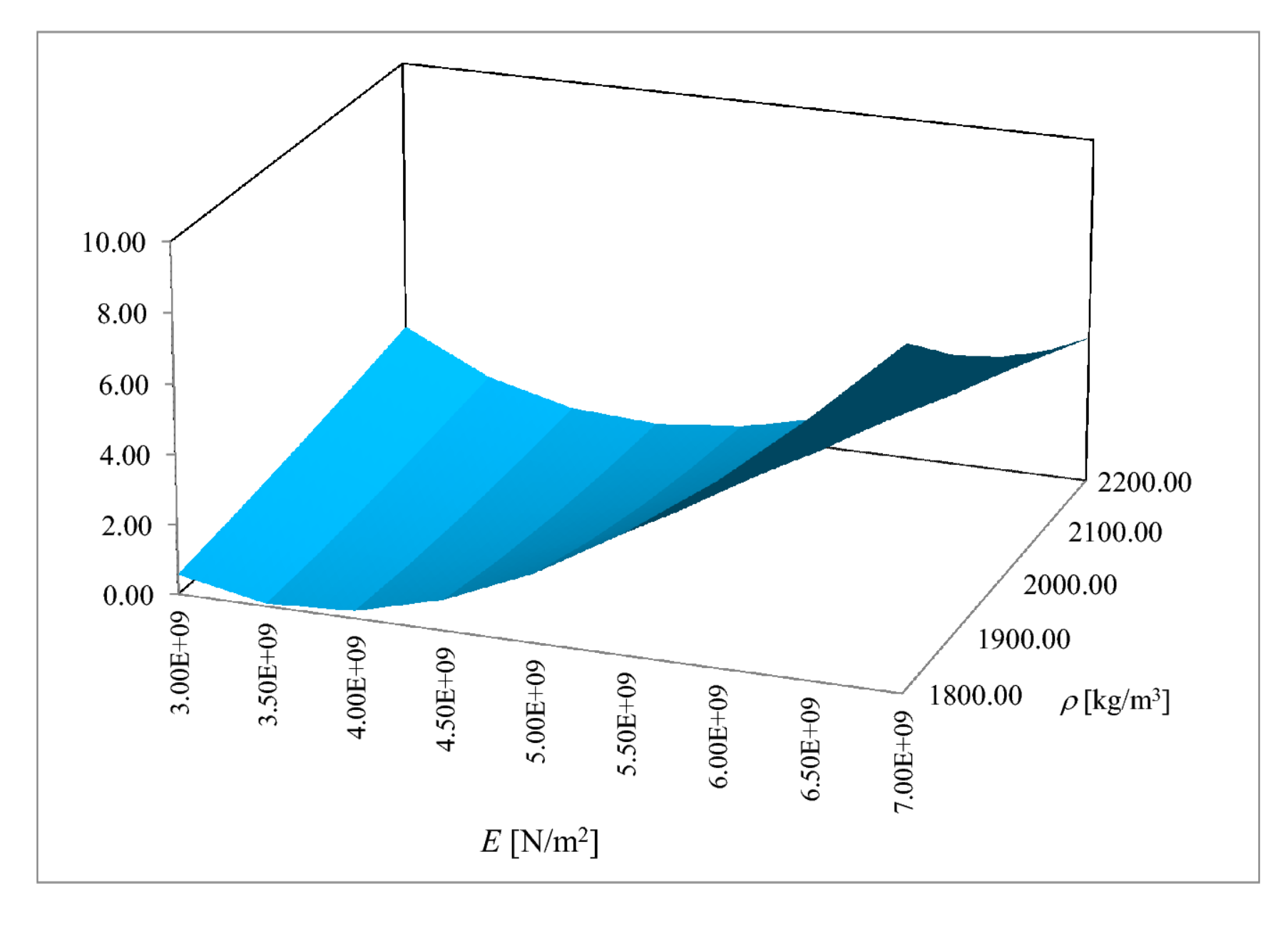}
\caption{Function $e^l(E, \rho)$: linear case.} \label{surf_lin}
\end{figure}

\begin{figure}\centering
\includegraphics[width=15cm]{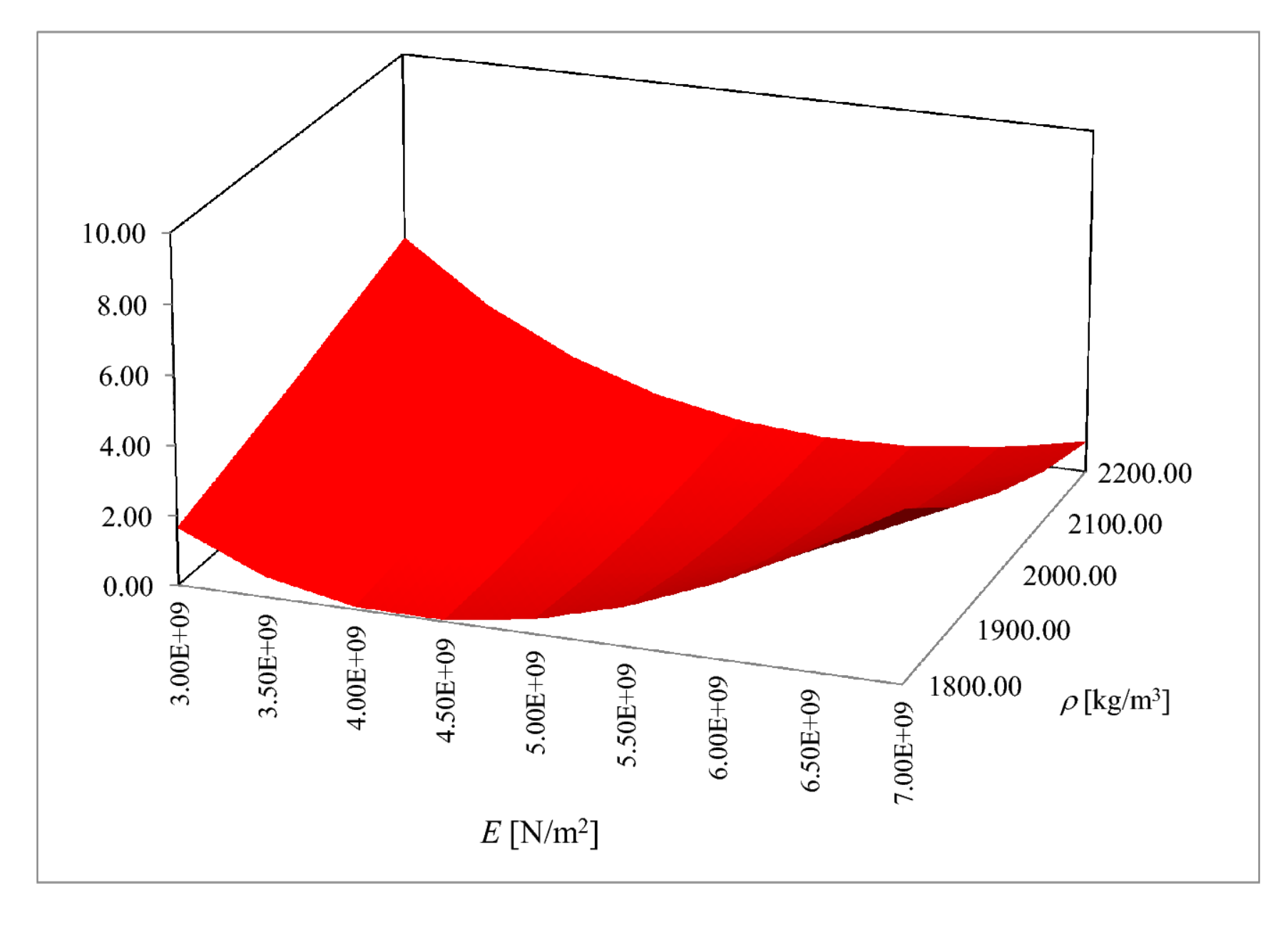}
\caption{Function $e(E, \rho)$: nonlinear case.} \label{surf_mas}
\end{figure}

\begin{figure}\centering
\includegraphics[width=10cm]{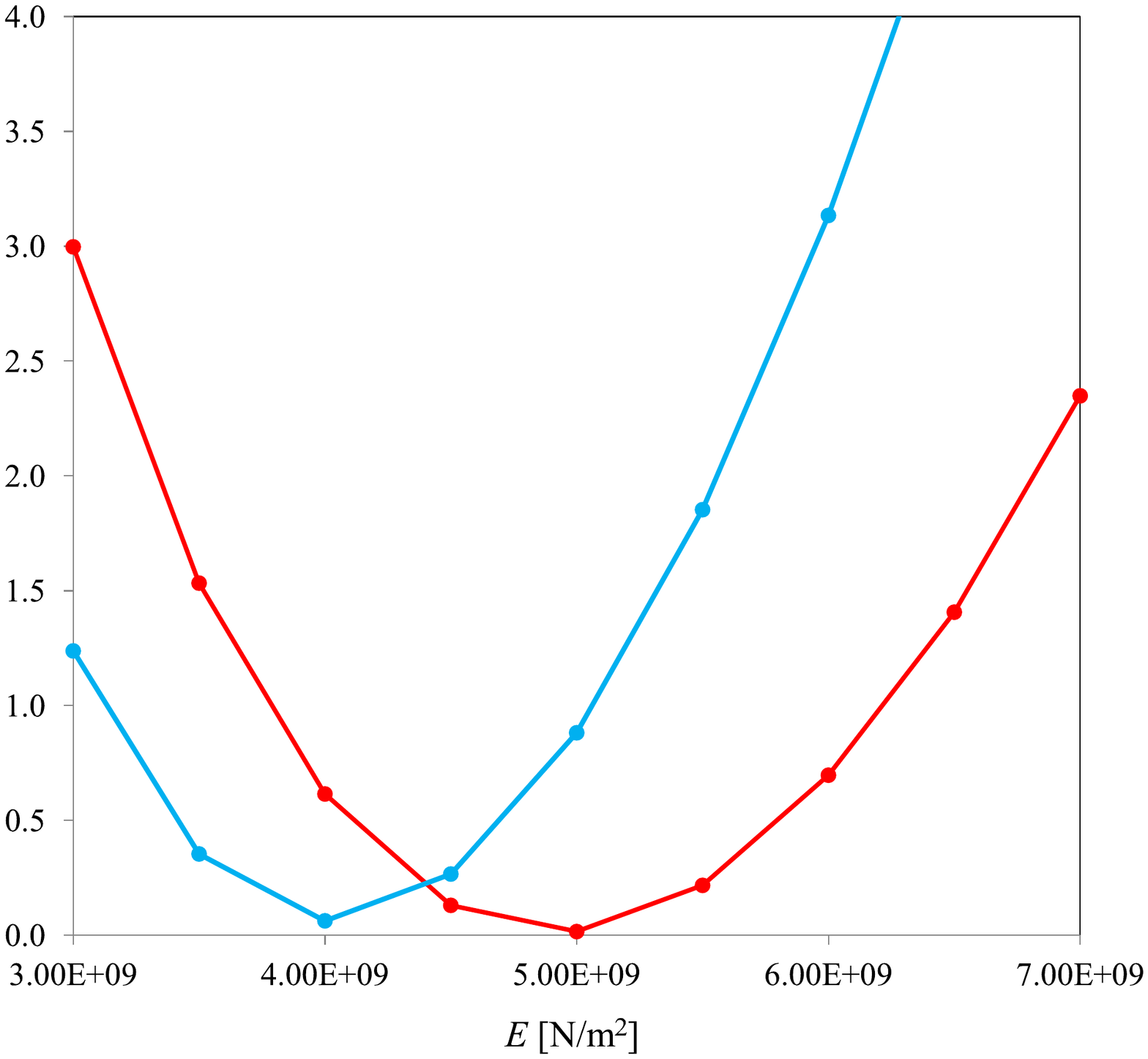}
\caption{Functions $e(E,\rho)$ (red line) and $e^l(E,\rho)$ (blue
line) vs. E for $\rho=\unit{2000}{\kilogram\per\metre^3}$.}
\label{sanfrediano_curves}
\end{figure}

\begin{figure}
\includegraphics[width=16cm]{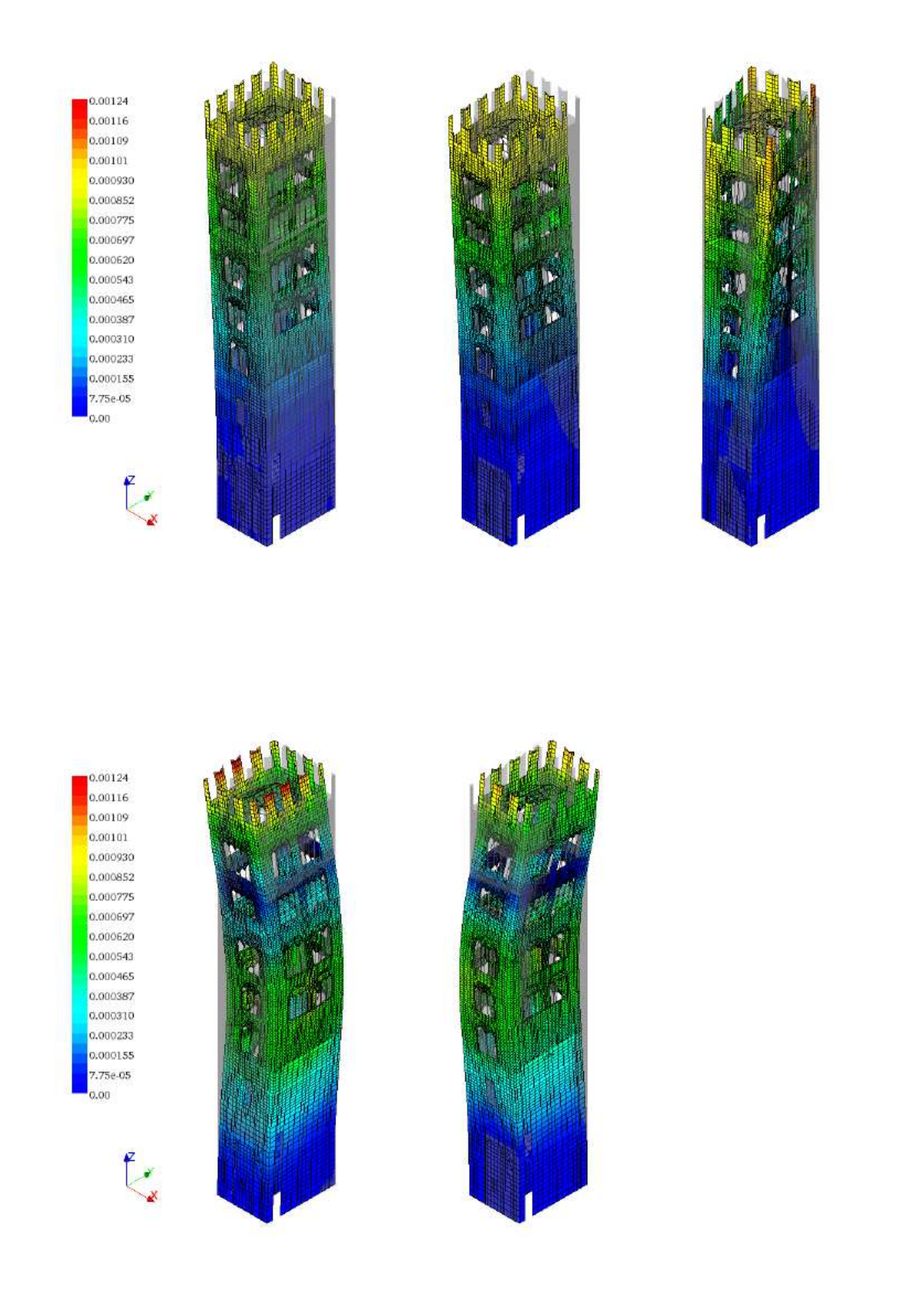}
\caption{Mode shapes $\phi_i$, $i=1,..,5$ of the San Frediano bell
tower in the nonlinear case, for $E=\unit{5\cdot
10^9\,}{\pascal},\,\rho=\unit{2000}{\kilogram\per\metre^3}$.}\label{sanfrediano_modi}
\end{figure}

\begin{figure}
\includegraphics[width=14cm]{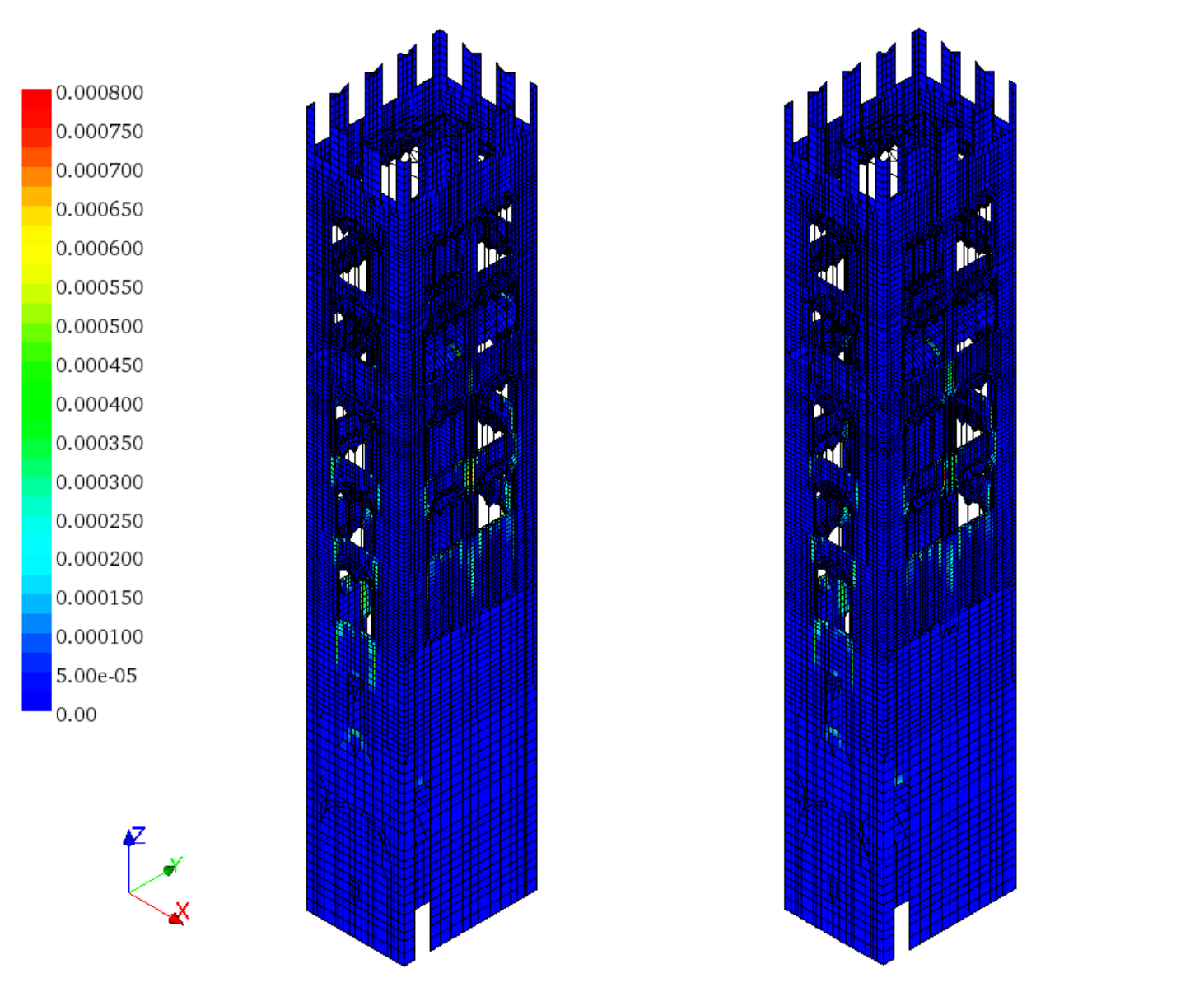}
\caption{Maximum principal fracture strains in the San Frediano bell
tower: inner (on the left) and outer (on the right) layers, for
$E=\unit{5\cdot
10^9\,}{\pascal},\,\rho=\unit{2000}{\kilogram\per\metre^3}$.}\label{sanfrediano_cracks}
\end{figure}

Figures \ref{surf_lin} and \ref{surf_mas} respectively show the functions $e^l$ and
$e$ vs. $E$ and $\rho$. They reach their
minimum values at $E=\unit{4\cdot 10^9\,}{\pascal}$ and
$\rho=\unit{2000}{\kilogram\per\metre^3}$ for the linear elastic
case and at  $E=\unit{5\cdot 10^9\,}{\pascal}$ and
$\rho=\unit{2000}{\kilogram\per\metre^3}$ for the nonlinear case.
Figure \ref{sanfrediano_curves} plots the two functions vs.
$E$ for $\rho=\unit{2000}{\kilogram\per\metre^3}$.  The first five
natural frequencies of the tower are shown in Table \ref{tab7},
where the experimental values $f_\text{i}^{\text{exp}}$ are compared
with the numerical values, in the linear elastic
$\left(f_\text{i}^{l}\right)$ and nonlinear
$\left(f_\text{i}\right)$ case. Columns $\Delta f _\text{i}^{l}$ and
$\Delta f_\text{i} $ show the differences between the numerical and
experimental values for the linear and nonlinear case,
respectively. The numerical frequencies differ from the experimental
values by no more than five percent. The nonlinear case is more
accurate in evaluating the higher modes. Figure
\ref{sanfrediano_modi} shows the first five mode shapes, calculated
in the nonlinear case for the optimal values $\big(E=\unit{5\cdot
10^9\,}{\pascal},\,\rho=\unit{2000}{\kilogram\per\metre^3}\big)$.
They are substantially equal to the linear elastic modes, which are
shown in \cite{sanfrediano}. The Figure clearly shows that the third
mode is torsional. Finally, Figure \ref{sanfrediano_cracks} shows the
maximum principal fracture strains for the tower in
equilibrium  with its own weight in the inner and outer layers of
the mesh for $E=\unit{5\cdot 10^9\,}{\pascal}$ and
$\rho=\unit{2000}{\kilogram\per\metre^3}$. The crack distribution
and the low values of the fracture strains indicate a modest level of
damage inside the structure: the fracture strains are concentrated
around the windows and lintels, and in the masonry supporting the
vault under the bell chamber. This confirms the observations made upon visual
inspection. Crack strains are concentrated in regions of the
structure quite far from those of the highest curvatures for the
first  mode shapes, which in fact remain very similar to those
calculated in the linear elastic case. However, due to the
nonlinearity of the constitutive equation adopted, the model
updating procedure yields higher values of Young's modulus,
which increases by about
the twenty--five percent with respect to the linear elastic case.


\pagebreak

%
%

 \listoffigures

\section{Conclusions} \label{sec5}

The paper deals with the modal analysis of masonry structures. In
particular, a new procedure for calculating the natural frequencies
and mode shapes of masonry buildings has been implemented within the
NOSA--ITACA code, to take into account the inability of masonry to
withstand tensile stresses. Some structures have been analyzed and
the influence on their dynamic properties of the fracture strains
induced by their self-weights and external loads investigated. In
particular, the new algorithm has been applied to the
finite--element model of the San Frediano bell tower in Lucca, and
the results of the numerical simulation compared with those from an
experimental campaign conducted on the tower in June, 2015. The new
procedure provides for more realistic simulations of the dynamic
behaviour of  masonry structures in the presence of cracks and, when
combined with experimental data, can be helpful in assessing damage.
Moreover, the dependence of the natural frequencies on external
loads (the case of thermal loads is under development) can be
assessed, and used to interpret data from long--term monitoring
protocols.

\def\baselinestretch{1}

\begin{table}[b]\centering
\begin{tabular}[l]{|c|c|c|c|c|c|c|c|c|c|}
\hline $\begin{array}{c}\text{Lateral}\\\text{load [N]}\end{array}$&
$\textbf{f}_\mathbf{1}[\hertz]$& \multicolumn{2}{|c|}{
$\begin{array}{c}\text{Eff. modal}\\\text{mass [\%]}\end{array}$}&
$\textbf{f}_\mathbf{2}[\hertz]$& \multicolumn{2}{|c|}{
$\begin{array}{c}\text{Eff. modal}\\\text{mass [\%]}\end{array}$}&
$\textbf{f}_\mathbf{3}[\hertz]$& \multicolumn{2}{|c|}{
$\begin{array}{c}\text{Eff. modal}\\\text{mass [\%]}\end{array}$} \\
\hline & &y & z &  &
y & z&  & y & z \\
\hline 0 & \textbf{6.47} &80.73 & 0.00 & \textbf{25.48} &
0.00 & 0.00& \textbf{53.79} & 0.00 & 81.06 \\
\hline 9000 &\textbf{6.44}&80.73&0.00& \textbf{25.48} &0.00&0.00&
\textbf{53.74} &0.00&80.76
\\\hline 10125 & \textbf{6.13} &80.32 & 0.00 & \textbf{25.27} &
0.00 & 0.00& \textbf{52.63} & 2.95 & 54.97 \\
\hline 11250 & \textbf{5.74}&79.98 & 0.00 & \textbf{24.36} &
0.00 & 0.00& \textbf{51.22} & 4.42& 44.12 \\
\hline 12375 & \textbf{5.23}&79.53 & 0.00 & \textbf{23.12} &
0.00 & 0.00& \textbf{48.97} & 6.05& 32.33 \\
\hline 13500 & \textbf{4.72}&79.09 & 0.00 & \textbf{21.80} &
0.00 & 0.00& \textbf{46.14} & 7.38& 22.13 \\
\hline 14625 & \textbf{4.21}&78.77 & 0.00 & \textbf{19.95} & 0.00 &
0.00& \textbf{43.75} &
8.10& 16.40 \\
\hline 15750 &\textbf{3.71}&78.35 & 0.00 & \textbf{18.34} &
0.00 & 0.00& \textbf{40.69} & 8.60& 12.06 \\
\hline
\end{tabular}

\vspace{1\baselineskip}

\begin{tabular}[l]{|c|c|c|c|c|c|c|c|c|c|}
\hline $\begin{array}{c}\text{Lateral}\\\text{load }
$[N]$\end{array}$& $\textbf{f}_\mathbf{4}[\hertz]$&
\multicolumn{2}{|c|}{ $\begin{array}{c}\text{Eff. modal}\\\text{mass
[\%]}\end{array}$}& $\textbf{f}_\mathbf{5}[\hertz]$&
\multicolumn{2}{|c|}{ $\begin{array}{c}\text{Eff. modal}\\\text{mass
[\%]}\end{array}$}& $\textbf{f}_\mathbf{6}[\hertz]$&
\multicolumn{2}{|c|}{
$\begin{array}{c}\text{Eff. modal}\\\text{mass [\%]}\end{array}$} \\
\hline & &y & z &  &
y & z&  & y & z \\
\hline 0 & \textbf{55.96} &8.77& 0.00 & \textbf{96.43} &
0.00 & 0.00& \textbf{145.30} & 3.07 & 0.00\\
\hline 9000 &\textbf{55.71}&8.77&0.00& \textbf{96.37} &0.00&0.00&
\textbf{144.79}
&3.07&0.00\\
\hline 10125 & \textbf{54.69} &6.15& 25.67& \textbf{94.51}&
0.00 & 0.00& \textbf{141.80} & 3.10 & 0.00 \\
\hline 11250 & \textbf{54.30}&4.90 & 36.06 & \textbf{90.15} &
0.00 & 0.00& \textbf{137.51} & 3.11& 0.00 \\
\hline 12375 & \textbf{53.46}&3.41& 47.18 & \textbf{85.85} &
0.00 & 0.00& \textbf{130.80}& 3.27& 0.00 \\
\hline 13500 & \textbf{52.31}&2.07&56.67& \textbf{82.42}&0.00&0.00&\textbf{122.50}&3.49&0.00 \\
\hline 14625 & \textbf{50.79}&1.22 & 61.54 & \textbf{76.97} & 0.00 &
0.00& \textbf{115.60} &
3.72& 0.00\\
\hline 15750 &\textbf{49.37}&0.06 & 64.93& \textbf{71.94} &
0.00 & 0.00& \textbf{108.10}&3.92& 0.00 \\
\hline
\end{tabular}
\caption{The beam's natural frequencies $\text{f}_\text{i}$,
$i=1,...,6$ and the effective modal masses of the corresponding mode
shapes versus load increments.} \label{tab1}
\end{table}

\begin{table}\centering
\begin{tabular}[c]{|c|c|c|c|c|c|c|}
\hline \multicolumn{7}{|c|}{MAC -- $M$}\\
\hline
&$\phi_1$& $\phi_2$  & $\phi_3$  & $\phi_4$  & $\phi_5$  & $\phi_6$ \\
\hline $\phi_1^{l}$&0.99&0.00&0.06&0.03&0.00&0.00\\
\hline $\phi_2^{l}$&0.00&0.99&0.02&0.04&0.12&0.00\\
\hline $\phi_3^{l}$&0.05&0.04&0.41&0.90&0.04&0.01\\
\hline $\phi_4^{l}$&0.04&0.00&0.88&0.41&0.01&0.19\\
\hline $\phi_5^{l}$&0.00&0.12&0.00&0.04&0.95&0.00\\
\hline $\phi_6^{l}$&0.00&0.00&0.16&0.09&0.00&0.93\\
\hline
\end{tabular}
\caption{MAC--$M(\phi_\text{i}^l,\phi_\text{j})$, for i,j = 1, 2,
..., 6 at the last load increment.} \label{tab2}
\end{table}

\begin{table}\centering
\begin{tabular}[c]{|c|c|c|c|c|c|c|c|c|c|}
\hline $\begin{array}{c}\text{Total}\\\text{vertical}\\\text{load
}[N]\end{array}$& $\textbf{f}_\mathbf{1}[\hertz]$&
\multicolumn{2}{|c|}{ $\begin{array}{c}\text{Eff. modal}\\\text{mass
[\%]}\end{array}$}& $\textbf{f}_\mathbf{2}[\hertz]$&
\multicolumn{2}{|c|}{ $\begin{array}{c}\text{Eff. modal}\\\text{mass
[\%]}\end{array}$}& $\textbf{f}_\mathbf{3}[\hertz]$&
\multicolumn{2}{|c|}{
$\begin{array}{c}\text{Eff. modal}\\\text{mass [\%]}\end{array}$} \\
\hline & &x & z &  &
x & z&  & x & z \\
\hline 0 & \textbf{6.54} &59.43 & 0.00 & \textbf{18.81} &
8.75 & 0.00& \textbf{22.49}& 0.00 & 7.92 \\
\hline 144605 &\textbf{6.53}&59.29&0.00& \textbf{18.50} &8.83&0.00&
\textbf{22.01} &0.00&6.98
\\\hline $\text{P}_1$=150605 & \textbf{5.98} &51.21 & 0.00 & \textbf{12.57} &
15.47 & 0.00& \textbf{21.04} & 0.00& 5.91 \\
\hline $\text{P}_2$=152605 & \textbf{4.79}&28.54 & 0.00 &
\textbf{9.00} &
37.27 & 0.00& \textbf{18.79} & 0.00& 2.92 \\
\hline $\text{P}_3$=153605 & \textbf{3.61}&15.38 & 0.00 &
\textbf{8.10} &
50.00 & 0.00& \textbf{15.43} & 0.00& 0.92 \\
\hline $\text{P}_4$=154605 & \textbf{2.54}&9.91 & 1.32&
\textbf{7.74} &
77.4 & 0.00& \textbf{12.06} & 0.00& 0.30 \\
\hline
\end{tabular}

\vspace{1\baselineskip}

\begin{tabular}[c]{|c|c|c|c|c|c|c|}
\hline $\begin{array}{c}\text{Total}\\\text{vertical}\\\text{load
}[N]\end{array}$& $\textbf{f}_\mathbf{4}[\hertz]$&
\multicolumn{2}{|c|}{ $\begin{array}{c}\text{Eff. modal}\\\text{mass
[\%]}\end{array}$}& $\textbf{f}_\mathbf{5}[\hertz]$&
\multicolumn{2}{|c|}{ $\begin{array}{c}\text{Eff. modal}\\\text{mass
[\%]}\end{array}$}
\\
\hline & &x & z &  &
x & z \\
\hline 0 & \textbf{33.74} &0.00& 1.57 & \textbf{45.37} &
11.88 & 0.00\\
\hline 144605 &\textbf{32.44}&0.00&2.82& \textbf{44.80} &11.28&0.00\\
\hline $\text{P}_1$=150605 & \textbf{30.34} &0.00 & 4.55 &
\textbf{38.48} &
6.74 & 0.00\\
\hline $\text{P}_2$=152605 & \textbf{27.28}&0.00 & 8.91 &
\textbf{32.70}&
3.96 & 0.00\\
\hline $\text{P}_3$=153605 & \textbf{25.45}&0.00 & 11.08 &
\textbf{28.19} &
2.98 & 0.00 \\
\hline $\text{P}_4$=154605 & \textbf{24.20}&0.00 & 11.70&
\textbf{23.60} &
1.39 & 0.00\\
\hline
\end{tabular}
\caption{The structure's natural frequencies $\text{f}_\text{i}$,
$i=1,..,5$ and the effective modal masses of the corresponding mode
shapes versus load increments. Load P at position
$X_4$.}\label{tab3}
\end{table}

\begin{table}\centering
\begin{tabular}[c]{|c|c|c|c|c|c|}
\hline \multicolumn{6}{|c|}{MAC -- $M$}\\
\hline
&$\phi_1$& $\phi_2$  & $\phi_3$  & $\phi_4$  & $\phi_5$ \\
\hline $\phi_1^{l}$&0.64&0.77&0.00&0.00&0.00\\
\hline $\phi_2^{l}$&0.71&0.59&0.05&0.08&0.31\\
\hline $\phi_3^{l}$&0.14&0.11&0.68&0.43&0.56\\
\hline $\phi_4^{l}$&0.07&0.06&0.52&0.63&0.07\\
\hline $\phi_5^{l}$&0.10&0.09&0.19&0.17&0.57\\
\hline
\end{tabular}
\caption{MAC--$M(\phi_\text{i}^l,\phi_\text{j})$, for i,j = 1, 2,
..., 5 at the last load increment. Load P at position
$X_4$.}\label{tab4}
\end{table}

\begin{table}\centering
\begin{tabular}[l]{|c|c|c|c|c|c|c|c|c|c|}
\hline $\begin{array}{c}\text{Total}\\\text{vertical}\\\text{load
}[N]\end{array}$& $\textbf{f}_\mathbf{1}[\hertz]$&
\multicolumn{2}{|c|}{ $\begin{array}{c}\text{Eff. modal}\\\text{mass
[\%]}\end{array}$}& $\textbf{f}_\mathbf{2}[\hertz]$&
\multicolumn{2}{|c|}{ $\begin{array}{c}\text{Eff. modal}\\\text{mass
[\%]}\end{array}$}& $\textbf{f}_\mathbf{3}[\hertz]$&
\multicolumn{2}{|c|}{
$\begin{array}{c}\text{Eff. modal}\\\text{mass [\%]}\end{array}$} \\
\hline & &x & z &  &
x & z&  & x & z \\
\hline 0 & \textbf{6.54} &59.43 & 0.00 & \textbf{18.81} &
8.75 & 0.00& \textbf{22.49}& 0.00 & 7.92 \\
\hline 144605 &\textbf{6.53}&59.29&0.00& \textbf{18.50} &8.83&0.00&
\textbf{22.01} &0.00&6.98
\\\hline $\text{P}_1$=150605 & \textbf{6.36} &57.80 & 0.00 & \textbf{17.09} &
9.99 & 0.00& \textbf{18.82} & 0.00& 0.70\\
\hline $\text{P}_2$=152605 & \textbf{6.19}&55.55 & 0.00 &
\textbf{15.06} &
11.89 & 0.00& \textbf{16.47} & 0.00& 0.00\\
\hline $\text{P}_3$=153605 & \textbf{6.06}&53.07 & 0.00 &
\textbf{13.41} &
14.08 & 0.00&\textbf{14.26} & 0.00& 0.60 \\
\hline $\text{P}_4$=154605 & \textbf{5.97}&51.06 & 0.00&
\textbf{12.44} &
15.92 & 0.00& \textbf{13.25} & 0.00& 0.10 \\
\hline
\end{tabular}

\vspace{1\baselineskip}

\begin{tabular}[l]{|c|c|c|c|c|c|c|}
\hline $\begin{array}{c}\text{Total}\\\text{vertical}\\\text{load
}[N]\end{array}$& $\textbf{f}_\mathbf{4}[\hertz]$&
\multicolumn{2}{|c|}{ $\begin{array}{c}\text{Eff. modal}\\\text{mass
[\%]}\end{array}$}& $\textbf{f}_\mathbf{5}[\hertz]$&
\multicolumn{2}{|c|}{ $\begin{array}{c}\text{Eff. modal}\\\text{mass
[\%]}\end{array}$}
\\
\hline & &x & z &  &
x & z \\
\hline 0 & \textbf{33.74} &0.00& 1.57 & \textbf{45.37} &
11.88 & 0.00\\
\hline 144605 &\textbf{32.44}&0.00&2.82& \textbf{44.80} &11.28&0.00\\
\hline $\text{P}_1$=150605 & \textbf{26.00} &0.00 & 12.76&
\textbf{40.18} &
0.00 & 5.55\\
\hline $\text{P}_2$=152605 & \textbf{24.37}&0.00 & 14.06 &
\textbf{39.31}&
0.00 & 3.69\\
\hline $\text{P}_3$=153605 & \textbf{23.34}&0.00 & 13.78 &
\textbf{37.13} &
2.74 & 0.00 \\
\hline $\text{P}_4$=154605 & \textbf{22.96}&0.00 & 13.41&
\textbf{35.52} &
2.09 & 0.00\\
\hline
\end{tabular}
 \caption{The structure's natural frequencies
$\text{f}_\text{i}$, $i=1,..,5$, and the effective modal masses of
the corresponding mode shapes versus load increments. Load P at
position $X_7$.} \label{tab5}
\end{table}

\begin{table}\centering
\begin{tabular}[c]{|c|c|c|c|c|c|}
\hline \multicolumn{6}{|c|}{MAC -- $M$}\\
\hline
&$\phi_1$& $\phi_2$  & $\phi_3$  & $\phi_4$  & $\phi_5$ \\
\hline $\phi_1^{l}$&0.98&0.17&0.00&0.00&0.00\\
\hline $\phi_2^{l}$&0.17&0.98&0.00&0.00&0.04\\
\hline $\phi_3^{l}$&0.00&0.00&0.46&0.89&0.00\\
\hline $\phi_4^{l}$&0.00&0.00&0.64&0.34&0.00\\
\hline $\phi_5^{l}$&0.00&0.03&0.00&0.00&0.75\\
\hline
\end{tabular}
 \caption{MAC--$M(\phi_\text{i}^l,\phi_\text{j})$, for i,j =
1, 2, ..., 5 at the last load increment. Load P at position $X_7$.}
\label{tab6}
\end{table}

\begin{table}\centering
\begin{tabular}[c]{|c|c|c|c|c|c|}
\hline
&$f_\text{i}^{\text{exp}}$& $f_\text{i}^{l}$  & $\Delta f _\text{i}^{l}$  & $f_\text{i}$  & $\Delta f_\text{i} $ \\
\hline Mode Shape 1&1.14&1.15&0.01&1.18&0.04\\
\hline Mode Shape 2&1.38&1.38&0.00&1.41&0.02\\
\hline Mode Shape 3&3.44&3.60&0.05&3.39&-0.01\\
\hline Mode Shape 4&4.60&4.41&-0.04&4.61&0.00\\
\hline Mode Shape 5&5.34&5.35&0.00&5.44&0.02\\
\hline
\end{tabular}
 \caption{Natural frequencies [Hz] of the tower: experimental values $f_\text{i}^\text{exp}$,
 numerical values $f_\text{i}^l$ identified on the linear elastic model; numerical values $f_\text{i}$ identified  on the masonry--like model. }
\label{tab7}
\end{table}

\end{document}